\documentclass[12pt]{article}

\usepackage{amsmath,amsfonts,amsthm,amssymb}
\usepackage{setspace}
\usepackage{fancyhdr}
\usepackage{lastpage}
\usepackage{extramarks}

\usepackage[moderate]{savetrees}

\usepackage{chngpage}
\usepackage{soul,color}
\usepackage{graphicx,float,wrapfig}
\usepackage{listings}
\usepackage{tikz}
\usepackage{harpoon}
\usepackage[ruled]{algorithm2e}
\usepackage{hyperref}
\usepackage{bm}
\usepackage{indentfirst}
\usepackage{natbib}
\usepackage{verbatim}
\usepackage{pgfplots}
\pgfplotsset{compat=1.18}

\hypersetup{
    colorlinks=true,
    citecolor=blue,
    linkcolor=blue,
    urlcolor=blue,
    bookmarksopen=true,
    pdfstartview={XYZ null null 1.00},
    pdfpagelayout={SinglePage}
}

\newtheorem{theorem}{Theorem}
\newtheorem{lemma}{Lemma}
\newtheorem{proposition}{Proposition}
\newtheorem{corollary}{Corollary}
\newtheorem{definition}{Definition}

\newtheorem{example}{Example}
\newtheorem{assumption}{Assumption}
\newtheorem{remark}{Remark}

\newtheorem{claim}{Claim}

\usetikzlibrary{arrows,graphs}
\newcommand{\enabstractname}{Abstract}

\newenvironment{enabstract}{%
  \noindent\mbox{}\hfill{\bfseries \enabstractname}\hfill\mbox{}\par
  \vskip 2.5ex}{\par\vskip 2.5ex}

\graphicspath{{figure/}}                                 %% Path of Figures
\usepackage[a4paper]{geometry}                           %% Paper size
\begin{document}
\title{Money Burning Improves Mediated Communication\thanks{We thank Nemanja Antic, Roberto Corrao, Yifan Dai, Piotr Dworczak, Françoise Forges, Alexander Jakobsen, Zhonghong Kuang, Xiao Lin, Elliot Lipnowski, Harry Pei, Fan Wu, Xingye Wu, Junjie Zhou, and the audience at WINE 2025 for their helpful comments and discussions. This paper was circulated with the title ``Mediated Information Design with Money Burning for Commitment Power." }}
\date{\today}
\author{Yi Liu\thanks{\baselineskip=0.7\normalbaselineskip
	Department of Economics, Yale University. {\tt yi.liu.yl2859@yale.edu}} \and Yang Yu\thanks{\baselineskip=0.7\normalbaselineskip
	School of Economics and Management, China University of Petroleum-Beijing. {\tt yangyu@cup.edu.cn}} }
\maketitle
\begin{enabstract}
Can wasteful money burning improve strategic communication? We show that it can, but only with intermediate commitment. In mediated communication with report-contingent burning, the mediator can use costly messages to discipline deviations and make persuasive messages credible. Under transparent motives, increasing the burning budget strictly raises the Sender's payoff once the budget is large enough, unless mediated communication with money burning collapses to cheap talk. With an unbounded budget, the value equals a robust Bayesian persuasion payoff, or equivalently the payoff of a cautious Sender. The framework clarifies commitment through smart contracts and Web 3.0 mediation.

\vspace{1ex}
\noindent \textbf{Keywords:} mediated communication, money burning, mechanism design, commitment
\noindent \textbf{JEL Classification:} D82, D83
\end{enabstract}

\newpage

\section{Introduction}\label{sec:intro}
When can a purely wasteful action improve the Sender's payoff in strategic communication? Money burning has long been viewed as a natural way for an informed Sender to make his messages more credible. Yet credibility and payoff improvement are distinct requirements. Burning can relax incentive constraints only by lowering the Sender's continuation payoff, so the instrument that makes communication credible also destroys part of the surplus that better communication is meant to create. A communication strategy may therefore become credible only because the Sender dissipates the entire gain from improved credibility, and possibly more. The central question is not merely whether money burning can support more informative communication, but whether it can raise the Sender's equilibrium payoff net of the resources destroyed.

This payoff question depends on how much commitment the Sender has. At the no-commitment extreme, money burning is only a voluntary costly signal. The cheap-talk literature shows that, outside knife-edge cases, costly messages make communication influential only when cheap talk without burning is already influential; they mainly refine existing credible communication rather than create it.\footnote{See \autoref{sec:ct_money_burning} for a discussion of money burning in cheap talk and its contrast with mediated money burning.} Under transparent motives, this limitation becomes a payoff limitation: a persuasive report can be made incentive compatible only by burning away the payoff gain that makes it attractive. Thus voluntary burning cannot strictly improve the Sender's payoff beyond what cheap talk already supports. At the full-commitment extreme, burning is redundant because the Sender can directly commit to an information structure, so any outcome with burning can be replicated without destroying resources. Money burning can therefore be valuable only with intermediate commitment: enough commitment to make burning enforceable, but not enough to eliminate reporting incentives. Only in this intermediate case does the joint design of messages and report-contingent burning determine whether credibility gains translate into payoff gains.

We study this intermediate environment through mediated communication. In the classical mediated communication (MD) protocol of \cite{myerson1982optimal,forges1986approach}, the Sender can commit to a mediator who maps his report into a message to the Receiver, but after his type is realized the Sender still chooses which report to send. Since then, most related studies have focused on characterizing MD and its corresponding equilibrium payoff. We ask a different question: once the Sender has enough commitment to delegate report-contingent consequences to a mediator, can he improve communication by committing the mediator to impose money burning based on reports?

The protocol we analyze, mediated communication with money burning (MDMB), augments MD in the minimal way needed to study this question. Before the type is realized, the Sender commits to a mediator who maps each report into both an output message to the Receiver and an observable amount of burned money. The mediator can enforce the consequences of a report, but the Sender still privately observes his type and chooses his report. Thus the Sender has enough commitment to make burning credible, but not enough commitment to eliminate reporting incentives. This is exactly the region in which money burning has a distinct economic role: it needs commitment to be enforceable, but it needs incomplete commitment to be useful.

We characterize MDMB under a budget constraint on money burning, which captures limited liability or bounded credit. Our analysis has two goals. First, we characterize the Sender's optimal equilibrium payoff and the corresponding optimal design. Second, and more importantly, we identify when money burning strictly improves mediated communication, or equivalently when the Sender's value is strictly increasing in the burning budget.\footnote{It is clear that when the budget for money burning increases, the Sender's payoff weakly increases.} The second question is the key to understanding whether money burning is merely an additional instrument or whether it genuinely expands communication efficiency.

The economic force behind our results is that mediated commitment allows money burning to separate the persuasive content of a message from its private deviation value. We work under transparent motives: the Sender's payoff depends only on the Receiver's action and not directly on his type. This assumption is substantive, but it captures many economic environments in which different types of the Sender agree on which Receiver actions are desirable (\cite{chakraborty2010persuasion,lipnowski2020cheap,lipnowski2022persuasion}). Under transparent motives, all Sender types rank reports in the same way whenever reports differ only in the Receiver action they induce. Hence, absent money burning, a report that induces a more favorable action is attractive to every type, and mediated communication must satisfy a strong type-indifference condition using information alone. Report-contingent burning relaxes exactly this constraint: a message can remain persuasive because it induces a favorable Receiver response, while its net payoff can be lowered enough to prevent all types from mimicking it. The key is that the loss need not be attached one-for-one to every persuasive message. Through the mediator, a controlled sacrifice at some reports can be linked to the credibility of other reports, so the Sender may gain from the improved information structure even after paying the burning cost.

This logic also sharpens the contrast with cheap talk. In cheap talk, burning can only be attached to the message that is voluntarily chosen and must itself deter mimicking. In MDMB, by contrast, burning is an enforceable consequence of a report chosen through a mediator. The Sender can therefore use burning as an incentive instrument inside a mediated mechanism: some messages are valuable not because they are themselves attractive, but because they discipline deviations and make other, more persuasive messages credible.

The first challenge is methodological. In a general MDMB, the burned money can itself contain information about the Sender's type and hence affect the Receiver's action. The classical revelation principle for mediated communication cannot be applied directly because part of the payoff-relevant outcome occurs before the Receiver acts and is also observed by her; applying the standard argument would violate the full-commitment condition emphasized by \cite{bester2001contracting}. We therefore develop a revelation principle tailored to MDMB. It decomposes any equilibrium outcome into a canonical process in which the Sender first induces a posterior belief through a signaling scheme and then the mediator burns a deterministic amount of money as a function of that posterior. In the canonical representation, burned money contains no additional information beyond the posterior itself. This decomposition yields a belief-based program with incentive-compatibility, obedience, Bayes-plausibility, and budget constraints, and complements the limited-commitment revelation principle of \cite{doval2022mechanism}.

The reduced program leads to a characterization of the Sender's maximum equilibrium payoff. Under transparent motives, incentive compatibility requires all Sender types to obtain the same interim payoff. The Lagrange multipliers on these type-indifference constraints form an affine combination of types and distort the Sender's payoff at each posterior. We call the resulting object the Sender's generalized subjective payoff function. \autoref{thm:SenderOPT_bounded} shows that the Sender's maximum payoff equals the minimum, over affine multipliers, of the concavification value of this generalized subjective payoff function.

This characterization also identifies how an optimal MDMB uses burning. For a posterior at which the distorted probability weight is positive, the Sender burns no money and induces the Receiver's best response that is most favorable to him. We call such posteriors the \emph{persuasion group}. For a posterior at which the distorted probability weight is negative, the Sender burns as much as possible and induces the Receiver's best response that is least favorable to him. We call such posteriors the \emph{credibility-gaining group}. The latter messages are not valuable because they are frequently used or directly persuasive; they are valuable because mediated commitment ties them, through incentive compatibility, to the persuasion group. In MDMB, credibility-gaining messages lower deviation values in one part of the mechanism and thereby allow the Sender to use more persuasive messages elsewhere without having their gains fully dissipated by burning. \autoref{prop:opt_MDMB} shows that an optimal MDMB is characterized by the joint choice of the persuasion group, the credibility-gaining group, and the associated signaling scheme.

Our main result shows that this force is generically strict. \autoref{thm:credibility_gaining_group} establishes that, under a generic condition on the Receiver's payoff functions, when the budget exceeds the range of the Sender's value function, increasing the budget strictly improves the Sender's payoff unless the Sender's payoff collapses to the cheap-talk benchmark. Equivalently, if additional burning capacity does not improve mediated communication, then mediated communication with money burning offers no payoff advantage over cheap talk. Building on this result, \autoref{prop:improve_CT_C} shows that when commitment has value, money burning strictly improves MD for almost all priors once the budget exceeds the threshold. Thus money burning is not simply a refinement of communication equilibria; in the mediated environment, it generically expands the Sender's equilibrium payoff set.

We also characterize the value of MDMB when the budget is unbounded. As the budget grows, the role of the credibility-gaining group becomes asymptotically small in probability but remains essential for incentive compatibility. The negative Lagrange multipliers converge to zero, and the limiting multipliers can be interpreted as the Sender's subjective beliefs. The value of MDMB is therefore the minimum among all concavification values of the Sender's subjective payoff functions (\autoref{thm:min_max}).\footnote{This refers to the convex combination, i.e., the expectation under the Sender's subjective beliefs, of the truth-adjusted welfare functions defined in \cite{doval2023persuasion}.} Equivalently, it is the payoff of a cautious Sender with full commitment power, who chooses a signaling scheme to maximize his worst interim payoff across types.\footnote{A cautious Sender is one who maximizes his minimum payoff across types (\cite{doval2021information}).} It also equals the Sender's payoff in Bayesian persuasion with heterogeneous priors under the worst Sender subjective prior (\cite{alonso2016bayesian}). These equivalences connect mediated money burning to robust Bayesian persuasion and provide a micro-foundation for max-min persuasion objectives.

In practice, this form of mediated commitment arises naturally in environments with transparent and automatically executed procedures. One example is the Web 3.0 economy, as proposed by \cite{drakopoulos2022blockchain}. For instance, \cite{shaker2021online} describe Web 3.0 financial companies that sell their products to consumers through smart contracts. These companies input risk-related information into smart contracts, which then generate risk assessment results for consumers according to predetermined and transparent algorithms. In this setting, the companies act as Senders, consumers are Receivers, and smart contracts serve as mediators. Gas fees or token transfers from the companies to consumers can be interpreted as forms of money burning. More broadly, transparent algorithms can both enforce the pre-committed message-generating protocol and impose observable costs. Therefore, communication protocol in Web 3.0 is an application of our mechanism.

Finally, we use the framework to revisit the value of commitment in Web 3.0 communities. Since transparent algorithms and smart contracts can implement MDMB-like communication, the relevant benchmark without full commitment is not always cheap talk. We define the gap between Bayesian persuasion and the value of MDMB as the \textit{refined value of commitment} in Web 3.0 communities. \autoref{prop:allequal} shows that this refined value is positive if and only if the conventional value of commitment is positive. Nevertheless, \autoref{coro:generic_better} implies that the refined value is generically smaller than the conventional value of commitment. Algorithmic mediation and enforceable burning therefore partially mitigate, but generally do not eliminate, the loss from the absence of full commitment.

\subsection{Related Literature}
Our work contributes to the literature on strategic communication under different levels of commitment power. When the Sender has no commitment, \cite{crawford1982strategic} introduce the cheap talk (CT) model. When the Sender has full commitment, \cite{kamenica2011bayesian} present the Bayesian persuasion (BP) model. Both benchmarks are important, but they leave open the role of instruments that are meaningful only between the two extremes: obtaining full commitment is difficult in many real-world contexts, while the unverifiable nature of cheap talk often limits communication. Related work studies intermediate forms of commitment. \cite{min2021bayesian,lipnowski2022persuasion} consider environments in which the Sender's commitment power has a Bernoulli distribution over full commitment and no commitment; \cite{lin2022credible} studies the case in which the Sender cannot commit to the message-generating process but can commit to the marginal distribution of types and messages. \cite{bergemann2019information} provide a unified perspective on information design, persuasion, and mediation. \cite{frechette2022rules} study the effect of communication rules and commitment experimentally. Most closely on the comparison of protocols, \cite{corrao2023communication} analyze communication under BP, MD, and CT, but they do not take money burning into account. Our analysis can instead be viewed as a comparison among MD protocols with different burning budgets, isolating how an enforceable costly instrument changes the power of mediated commitment.

Within this broad literature, our paper is closest to work on mediated communication. Previous research, such as \cite{salamanca2021value}, characterizes the Sender's optimal equilibrium payoff under MD without money burning. Several studies, including \cite{goltsman2009mediation,ivanov2014beneficial}, identify optimal mediation plans for the Receiver, and \cite{ivanov2014beneficial} compares mediated communication with cheap talk. \cite{drakopoulos2022blockchain} establish a blockchain system as a mediator and show that designing costly messages can improve MD under transparent motives. We share their interest in mediated environments with costly messages, but our focus is different: we characterize the Sender's optimal communication efficiency in general and identify when costly messages, through money burning, strictly improve MD. Thus the main contribution of the paper is to extend the domain of mediated communication by incorporating money-burning tactics and studying the payoff effect of the burning budget.

Our paper is also related to the literature on communication with transfers and costly signals. Several studies analyze cheap talk or signaling games with money burning \cite{austen2000cheap,kartik2007note,karamychev2017optimal,gersbach2004money} and show that money burning generally cannot enhance the credibility of cheap talk. In the cheap-talk benchmark, we also find that money burning cannot improve the Sender's payoff under state-independent preferences. In cheap talk and signaling games, money burning can expand the set of equilibria and sustain more informative equilibria. However, a key limitation is that money burning can expand equilibria only when cheap talk itself is already influential. Moreover, these studies focus mainly on constructing such equilibria and do not study the payoff implications of money burning for the Sender. In contrast, we show that money burning can enhance the Sender's credibility, expand the equilibrium payoff set, and improve communication efficiency even when cheap talk alone is not influential. Relatedly, \cite{koessler2026belief} develop a belief-based characterization of costly signaling games using Bayes-plausible belief distributions and incentive-compatible interim values, and their results also support the limited role of costly signals in cheap talk under transparent motives; our belief-based program uses similar objects, but the burning amount in our model is an enforceable consequence of a mediated report rather than a voluntarily chosen signal. \cite{kolotilin2021relational} investigate monetary transfers in repeated communication, while \cite{sadakane2023multistage} examines repeated cheap talk games with monetary transfers from the Receiver to the Sender and shows that the equilibrium set can be larger than in the original long-term cheap talk setting. \cite{corrao2023mediation} analyzes mediation markets and characterizes the information and market outcomes of the revenue-maximizing mediator and the Sender-optimal mediator. Several papers study Bayesian persuasion with transferable utility or costly information, such as \cite{li2017discriminatory,bergemann2018design}. \cite{perez2022test} investigates Receiver-optimal experiment design when the Sender can costly falsify his private type.

Another important category of related work is mechanism design with limited commitment. \cite{wu2023eliciting} examines implementation in general outcome-contingent settings, which is more general than ours. \cite{bester2001contracting} show that the revelation principle can fail in limited-commitment environments, where the principal cannot fully commit to the outcome induced by the mechanism. \cite{doval2022mechanism} provide a general revelation principle for limited-commitment mechanism design. However, they require randomization over the mechanism-design part given the information realization. In our revelation principle, we further show that the money-burning part can be deterministic given the information realization.

\section{Model}\label{sec:model}

In \autoref{sec:model_pre}, we develop the basic model of the Sender-Receiver game, and in \autoref{sec:simplfy}, we introduce the methodology for simplifying the Sender's programming problem.

\subsection{Basic Setup}\label{sec:model_pre}

\paragraph{Primitives.} The basic game consists of two players: the Sender (he) and the Receiver (she). The Sender has the private information $\theta$, which denotes his type and belongs to a finite set $\Theta$. Type $\theta$ is drawn according to a full-support prior distribution $\mu_0\in\Delta(\Theta)$, which is common knowledge. The Receiver can choose an action $a$ from a finite set $A$. The Sender's value function only depends on the Receiver's action, which is denoted as $v(\cdot):A\rightarrow \mathbb{R}$; the Receiver's value function is $u(\cdot,\cdot):A\times\Theta\rightarrow \mathbb{R}$. 

\paragraph{Communication with money-burning mechanisms.} Before the game, the Sender can commit to a mediated communication with money-burning mechanism (or MDMB), which consists of an input set $M$, an output message set $S$, and a corresponding mechanism $\phi: M \rightarrow \Delta(S \times T)$, where $T \triangleq [0, C]$ and $C \in \mathbb{R}_{\geq 0} \cup \{+\infty\}$ denotes the Sender's exogenously given total budget. The MDMB prescribes how the Sender designs the message and determines the money-burning amount based on his private input. We restrict attention to the case where $M$, $S$, and the support of $\phi$ are all finite.	

\paragraph{The Sender-Receiver game.} At the beginning of the game, the Sender commits to the MDMB $(M,S,\phi)$ with a mediator. Then, the Sender's type $\theta$ is realized, and the Sender sends an input message $m\in M$ to the mediator. With probability $\phi(s,t|m)$, the mediator sends an output message $s\in S$ to the Receiver and burns $t\in T$ money from the Sender's account. After observing the money burned by the Sender and the output message $s$, the Receiver updates her belief and chooses an action $a\in A$. The ex-post payoffs of the Sender and the Receiver are $v(a)-t$ and $u(a,\theta)$, respectively.

\begin{remark}
    \emph{In our model, the Receiver observes the money burning. Relative to models with unobservable money burning, our framework requires less commitment from the Sender and applies to a broader range of environments, particularly when the mediator is replaced by transparent algorithms or when the burned money is provided as a subsidy to the Receiver.}
\end{remark}

After the commitment of MDMB $(M,S,\phi)$, the sub-game is denoted as $\mathcal{G}_{(M,S,\phi)}$. 

\paragraph{Beliefs and strategies.} The Sender's strategy in $\mathcal{G}_{(M,S,\phi)}$ prescribes a transition probability $\sigma: \Theta\rightarrow \Delta(M)$. As for the Receiver, the output message $s$ and the money burning amount $t$ together form the information set. For each information set $(s,t)$, the Receiver's strategy prescribes a transition probability $\alpha:S\times T\rightarrow \Delta(A)$. In each information set $(s,t)$, the Receiver must form a belief $\mu: S\times T \rightarrow \Delta(\Theta)$. We call the triple $(\sigma,\alpha,\mu)$ an assessment.

\paragraph{Equilibrium.} In this paper, we use Perfect Bayesian equilibrium (henceforth, PBE) as the solution concept of game $\mathcal{G}_{(M,S,\phi)}$. We denote the set of PBE of game $\mathcal{G}_{(M,S,\phi)}$ as $\mathcal{E}[\mathcal{G}_{(M,S,\phi)}]$. Formally, $(\sigma^*,\alpha^*,\mu^*)\in \mathcal{E}[\mathcal{G}_{(M,S,\phi)}]$ if the following three conditions hold:

Sender's optimality: for any $\theta\in \Theta$,
\begin{equation}\label{eqn_SenderOPT}
    \sigma^*(\theta)\in\arg\max_{\sigma(\theta)\in\Delta(M)}\sum_{m\in M,s\in S,t\in T,a\in A}\sigma(m|\theta)\phi(s,t|m)\alpha^*(a|s,t) (v(a)-t).
\end{equation}

Receiver's optimality: for any $s\in S,t\in T$,
\begin{equation}\label{eqn_ReceiverOPT}
\alpha^*(s,t)\in\arg\max_{\alpha(s,t)\in\Delta(A)}\sum_{\theta\in\Theta,a\in A}\mu^*(\theta|s,t)\alpha(a|s,t) u(a,\theta).
\end{equation}

Bayesian updating: for any $s\in S,t\in  T$ and $\theta\in\Theta$,
\begin{equation}\label{eqn_Bayesian}
\mu^*(\theta|s,t)\sum_{\theta'\in\Theta,m\in M}\mu_0(\theta')\sigma^*(m|\theta')\phi(s,t|m)=\mu_0(\theta)\sum_{m\in M}\sigma^*(m|\theta)\phi(s,t|m).
\end{equation}

\paragraph{Communication efficiency.} The Sender seeks to maximize his ex ante expected payoff across all possible PBEs. We can formulate the Sender's optimization problem as follows:
\begin{equation}\label{eqn:valueMDMB}
    \begin{split}
        \sup_{M,S,\phi} & \sum_{\theta\in\Theta}\mu_0(\theta) \sum_{m\in M,s\in S,t\in T,a\in A}\sigma^*(m|\theta)\phi(s,t|m)\alpha^*(a|s,t) (v(a)-t) \\
        s.t. &
        (\sigma^*,\alpha^*,\mu^*)\in \mathcal{E}[\mathcal{G}_{(M,S,\phi)}].
    \end{split}
\end{equation}
We focus on the effects of the prior $\mu_0$ and the budget $C$ on the Sender's maximum equilibrium payoff. Thus, we denote the value of (\ref{eqn:valueMDMB}) by $\mathcal{V}_{C}^*(\mu_0)$. When $C=0$, $\mathcal{V}^*_{0}(\mu_0)$ represents \emph{the value of MD}, which is the Sender's maximum payoff from standard mediated communication (MD) with no money burning; while when $C=+\infty$, we simply write $\mathcal{V}^*(\mu_0)$, representing \emph{the value of MDMB}.

\subsection{Simplifying the Problem}\label{sec:simplfy}
The revelation principle of \cite{myerson1982optimal,forges1986approach} cannot be applied directly to simplify the optimization problem (\ref{eqn:valueMDMB}). In their framework, payoff-relevant outcomes are determined only after information transmission. In our model, however, one payoff-relevant outcome, namely money burning, is observed before the Receiver acts and can affect her subsequent action through the information it conveys. Applying the classical revelation principle would therefore violate the full-commitment assumption \cite{bester2001contracting}. Hence, in this section, we develop a new technique to simplify the problem. Inspired by \cite{doval2022mechanism}, we apply the method of canonical mechanisms and canonical assessments to develop a new revelation principle for MDMB.

Here, we first formally define the canonical MDMBs and canonical assessments.
\begin{definition}[Canonical MDMBs]\label{def:can_MDMB}
    An MDMB is canonical if $M=\Theta$, $S=\Delta(\Theta)$, and there exists a signaling scheme $\pi:\Theta\rightarrow\Delta(\Delta(\Theta))$ and a deterministic function $x:\Delta(\Theta)\rightarrow  T$ such that $\pi$ satisfies the Bayesian updating condition\footnote{$\nu(\theta)\sum_{\theta'\in\Theta}\mu_0(\theta')\pi(\nu|\theta')=\mu_0(\theta)\pi(\nu|\theta)$ for all $\nu\in \Delta(\Theta)$.} and $\phi(\nu,x(\nu)|\theta)=\pi(\nu|\theta)$ for all $\theta\in\Theta$ and $\nu\in\textbf{supp}\{\pi(\theta)\}$.
\end{definition}

In canonical MDMBs, the input sets are the type sets, while the output sets are the sets of distributions of types. Although the canonical message space is written as $\Delta(\Theta)$, only finitely many posterior messages are used with positive probability in any canonical MDMB considered here. The output message of the canonical MDMB contains \emph{all} information transmitted to the Receiver, and the amount of money burning does not provide any additional information. Hence, $\phi$ in a canonical MDMB can be decomposed into two parts. The first part is a \emph{signaling scheme} $\pi$, and the second part is a \emph{money-burning scheme} $x$ that is contingent on the output message. This decomposition has a similar structure to the revelation principle in \cite{doval2022mechanism}, except that $x$ is a deterministic function. Henceforth, we use $(\pi,x)$ to refer to a canonical MDMB.

In a canonical MDMB, the canonical assessment ensures that the Sender's strategy is truthful-telling and the Receiver's posterior belief coincides with the output message.
\begin{definition}[Canonical assessments]\label{def:can_ass}
    For a canonical MDMB, an assessment $(\sigma,\alpha,\mu)$ is canonical if $\sigma(\theta|\theta)=1$ and $\mu(\nu,x(\nu))=\nu$ for any $\nu\in\textbf{supp}\{\pi(\theta)\}$.
\end{definition}

The following proposition shows that every MDMB has a corresponding canonical MDMB that maintains the same expected payoffs of the Sender and the Receiver. This proposition allows us to focus exclusively on the canonical MDMBs and the associated canonical assessment without loss of generality, as depicted in \autoref{fig:Revelation_Principle}. 

\begin{proposition}\label{prop:revelation_principle}
    For any MDMB $(M,S,\phi)$ and $(\sigma,\alpha,\mu)\in \mathcal{E}[\mathcal{G}_{(M,S,\phi)}]$, there exists a canonical MDMB $(\pi,x)$ and a canonical assessment $(\sigma^*,\alpha^*,\mu^*)\in \mathcal{E}[\mathcal{G}_{(\pi,x)}]$ such that the two assessments are payoff-equivalent.\footnote{Two assessments are payoff-equivalent if they induce the same ex ante distribution of payoffs for the Sender and the Receiver.}
\end{proposition}

\begin{figure}[htbp]
            \centering
            \begin{minipage}{0.45\linewidth}
            \begin{tikzpicture}[scale=2]
                \node at (-1.5,0){Sender};
                \node at (-1.5,-0.1)[below]{$\theta$};
            \node at (-1,0){$M$};
            \draw[->](-0.8,0)--(-0.4,0);
            \node at (0.1,0){$S\times T$};
            \node at (0.9,0){Receiver};
            \node at (-0.6,0.2){$\phi(\cdot|m)$};
            \end{tikzpicture}
            \end{minipage}
            \begin{minipage}{0.05\linewidth}
            $\Leftrightarrow$
            \end{minipage}
            \begin{minipage}{0.45\linewidth}
                \begin{tikzpicture}[scale=2]
                \node at (-1.6,0){Sender};
                \node at (-1.5,-0.1)[below]{$\theta$};
            \node at (-1.1,0){$\Theta$};
            \draw[->](-1,0)--(-0.6,0);
            \node at (-0.3,0){$\Delta(\Theta)$};
            \draw[->](-0.05,0)--(0.3,0)node[right]{$T$};
           \node at (0.1,0.2){$x(\cdot)$};
           \node at (1.15,0){Receiver};
            \node at (-0.8,0.2){$\pi(\cdot|\theta)$};
            \end{tikzpicture}
            \end{minipage}
            \caption{Revelation Principle}
            \label{fig:Revelation_Principle}
\end{figure}

The intuition of \autoref{prop:revelation_principle} is that for any MDMB $(M, S, \phi)$ with its corresponding PBE assessment $(\sigma, \alpha, \mu)$, each pair $(s, t) \in S \times T$ can be viewed as a new signal $s' \in S'$, where $S' = S \times T$ and $s' = (s, t)$, with the associated money burning defined as $x(s') = t$. Thus, we have $\mu(s, t) = \mu(s')$, leading to a decomposition structure. Furthermore, since the Receiver's actions depend on beliefs, each signal $s'$ can be replaced with the posterior belief $\mu(s')$, allowing us to substitute the signal space $S'$ with the belief space $\Delta(\Theta)$. The remaining challenge is handling cases where different signals $s'$ correspond to the same posterior belief but involve different burning amounts and different Receiver mixed actions. As shown in the appendix, by defining the money burning at each posterior $\mu(s')$ as the expected total payment conditional on $\mu(s')$, the payoff-equivalence property is maintained. The Receiver's action can be averaged in the same conditional way, and obedience is preserved because the best-response set is convex at a fixed posterior. This logic is general, and the result in \autoref{prop:revelation_principle} extends beyond state-independent preferences by the same argument. However, the subsequent analysis in this paper relies on the assumption of state-independent preferences, and we therefore maintain this assumption throughout.

A central feature of \autoref{prop:revelation_principle} is its decomposition structure, which facilitates the calculation of the value in (\ref{eqn:valueMDMB}). Moreover, interpreting the signal set as a set of posterior beliefs not only simplifies the notation but also lays the foundation for a belief-based framework. In the rest of this section, we explain how to apply \autoref{prop:revelation_principle} to simplify the optimization problem via the belief-based approach. This approach considers the ex ante distribution over Receiver's posterior beliefs, $p\in\Delta(\Delta(\Theta))$. By \cite{kamenica2011bayesian}, the Bayesian updating constraint is equivalent to requiring $p\in BP(\mu_0)$ where \( BP(\mu_0)\triangleq \big\{p\in\Delta(\Delta(\Theta))\big|\int_{\mu}\mu dp(\mu)=\mu_0\big\}.\) In other words, there is a mapping between signaling scheme $\pi$ and $p\in BP(\mu_0)$ through Bayesian updating. Therefore, we also call $p\in BP(\mu_0)$ the signaling scheme. 

For any posterior belief $\mu\in\Delta(\Theta)$, the Receiver's best response can be summarized in 
$
RO(\mu)\triangleq\big\{\alpha\in\Delta(A)\big|\textbf{supp}\{\alpha\}\subset \arg\max_{a'\in A}\sum_{\theta\in \Theta}\mu(\theta)u(a',\theta)\big\}.
$
Therefore, in PBEs, the Sender's value of the Receiver's action can be summarized in the Sender's \emph{belief-value correspondence}, $\mathbb{V}:\Delta(\Theta)\rightrightarrows \mathbb{R}$, where 
$
\mathbb{V}(\mu)\triangleq\big\{q\big|\exists \alpha\in RO(\mu),q=\sum_{a\in A}\alpha(a)v(a)\big\}.
$
Hence, the Receiver's optimality constraints can be converted to obedience constraints that $V(\mu)\in\mathbb{V}(\mu)$ where $V(\mu)$ denotes the Sender's value from the Receiver's action at posterior $\mu$.

Regarding the Sender's optimality constraints, suppose the canonical MDMB $(\pi, x)$ induces a distribution over posteriors denoted by $p$. A Bayes-plausible distribution $p$ over posteriors can be canonically implemented by defining the conditional distribution over posteriors given type $\theta$ through the Radon-Nikodym derivative $\frac{\mu(\theta)}{\mu_0(\theta)}$ with respect to $p$.\footnote{Formally, for any measurable $B\subseteq\Delta(\Theta)$, $\pi(B|\theta)=\int_{B}\frac{\mu(\theta)}{\mu_0(\theta)}\,dp(\mu)$. This is well defined because $\int_{\Delta(\Theta)}\frac{\mu(\theta)}{\mu_0(\theta)}\,dp(\mu)=1$ by Bayes plausibility.} Hence, the Sender's expected payoff when reporting type $\theta$ is given by $\int_{\mu} \frac{\mu(\theta)}{\mu_0(\theta)}(V(\mu) - x(\mu))\,dp(\mu).$
Because the Sender's payoff is state independent, the payoff from any report depends on the report-induced distribution over outcomes, not on the Sender's true type directly. Thus, truthful reporting is optimal for all types only if all type reports induce the same interim payoff. Therefore, the Sender's optimality condition requires type-indifference: that is, there exists a constant $k \in \mathbb{R}$ such that $
\int_{\mu} \frac{\mu(\theta)}{\mu_0(\theta)}(V(\mu) - x(\mu))\,dp(\mu) = k$ for all $\theta \in \Theta$.
We conclude with the following corollary.

\begin{corollary}\label{coro:simply_value_opt}
$\mathcal{V}^*_{C}(\mu_0)$ can be calculated using the following optimization problem.
\begin{alignat}{2}
    \max \quad & k & {} & \label{eqn:SenderOPT_Canonical}\\
    \mbox{s.t} \quad & k=\int_{\mu}\frac{\mu(\theta)}{\mu_0(\theta)}(V(\mu)-x(\mu))dp(\mu)& \quad & \forall \theta\in \Theta \tag{IC}\label{eqn:IC} \\
    & p\in BP(\mu_0) &\quad &  \tag{BP}\label{eqn:BP} \\
    & V(\mu) \in \mathbb{V}(\mu) & \quad & \forall \mu\in \Delta(\Theta)\tag{O}\label{eqn:Obe} \\
    & 0\leq x(\mu)\leq C &\quad & \forall \mu\in \Delta(\Theta) \tag{Bgt}\label{eqn:Bgt}
\end{alignat}
\end{corollary}

In this paper, we will use the following salesman's example to illustrate our main characterizations and intuitions. In this example, the Sender cannot benefit from cheap talk or mediated communication.
\begin{example}\label{example:illustrative}
    Consider a salesman problem between a consumer and a salesman. The consumer decides whether to purchase a product, whose quality is either high ($\theta^{H}$) or low ($\theta^{L}$). The salesman has private information about the true quality of the product, while the consumer has a prior belief that the product is high-quality with probability $0<\mu_0<\frac{1}{2}$. The market price of the product is fixed at $5$. The consumer's gross value from the product is $10$ when the quality is high and $0$ when the quality is low; hence, her net payoff from purchasing is $5$ under high quality and $-5$ under low quality, while her payoff from not purchasing is $0$. The salesman's payoff is determined by the consumer's decision: he receives a reward of $1$ from the producer as his commission for selling the product and receives nothing otherwise.
\end{example}

\section{Sender-optimal Equilibrium Payoffs}\label{sec:upper_bound}
In this section, we characterize the Sender's maximum equilibrium payoff $\mathcal{V}^*_{C}(\mu_0)$ for a fixed burning budget $C$ and explain the economics of optimal MDMB. The characterization identifies how an optimal mechanism splits posterior messages into a persuasion role and a credibility-gaining role. This structure will be used in the next section to study when increasing the burning budget strictly improves mediated communication.

For any Lagrange multiplier $\lambda\in \text{aff}(\Theta)$, define the distorted weight of posterior $\mu$ as
\[
    \omega_{\lambda}(\mu)\triangleq \sum_{\theta\in\Theta}\lambda(\theta)\frac{\mu(\theta)}{\mu_0(\theta)}.
\]
The multiplier $\lambda$ comes from the type-indifference constraints \eqref{eqn:IC}. Thus, $\omega_{\lambda}(\mu)$ measures how the shadow value of relaxing those constraints loads on posterior $\mu$. Since $\lambda$ belongs to the affine hull of $\Theta$, $\omega_{\lambda}(\mu)$ can be positive or negative. A positive distorted weight means that the Sender wants the posterior to generate the highest possible action payoff and no burning. A negative distorted weight means that the posterior is useful for disciplining deviations, so the Sender chooses the lowest possible action payoff and the largest burning amount.

We define the Sender's \emph{generalized subjective payoff function} at posterior $\mu\in\Delta(\Theta)$, denoted by $\hat{V}_{\lambda,C}(\mu)$, as follows
\begin{equation}\label{eqn:hatV_cost}
\hat{V}_{\lambda,C}(\mu)\triangleq\max\{\omega_{\lambda}(\mu)\max \mathbb{V}(\mu),\omega_{\lambda}(\mu) (\min \mathbb{V}(\mu)-C)\}.
\end{equation}
In addition, let $cav(f)$ denote the concave envelope of the function $f$. The characterization of the Sender's maximum equilibrium payoff is based on the generalized subjective payoff function and the concave envelope.
\begin{theorem}\label{thm:SenderOPT_bounded}
    $$\mathcal{V}^*_{C}(\mu_0)=\min_{\lambda\in \text{aff}(\Theta)}cav(\hat{V}_{\lambda,C})(\mu_0).$$
\end{theorem}

To compute the Sender's maximum equilibrium payoff, we conduct the two-step optimization approach. First, given any signaling scheme $p\in BP(\mu_0)$, we calculate the optimal money-burning scheme and the Receiver's response $V(\mu)$. Note that at this stage, the variables $x(\mu)$, $V(\mu)$, and $k$ are linear in the objective and constraints. By introducing the Lagrange  multiplier of (\ref{eqn:IC}) $\lambda$, the Lagrangian function is 
\[
k+\sum_{\theta\in\Theta}\lambda(\theta)\left[\int_{\mu}\frac{\mu(\theta)}{\mu_0(\theta)}(V(\mu)-x(\mu)) dp(\mu) - k\right].
\]
Thus, in the optimal solution, $(V(\mu),x(\mu))$ can be either $(\max \mathbb{V}(\mu),0)$ or $(\min \mathbb{V}(\mu),C)$. We know that $x(\mu)=0$ and $V(\mu)=\max\mathbb{V}(\mu)$ when $\omega_{\lambda}(\mu)>0$; $x(\mu)=C$ and $V(\mu)=\min\mathbb{V}(\mu)$ when $\omega_{\lambda}(\mu)< 0$. In addition, by the first-order condition of variable $k$, the Lagrange multiplier must satisfy that $\lambda\in \text{aff}(\Theta)$. Therefore, after optimizing over the money burning scheme and the Receiver's response, the objective becomes
\[
\min_{\lambda\in \text{aff}(\Theta)}\int_{\mu}\max\{\omega_{\lambda}(\mu)\max \mathbb{V}(\mu),\omega_{\lambda}(\mu)(\min \mathbb{V}(\mu)-C)\}dp(\mu)
\]
and the Sender aims to maximize the objective over $p\in BP(\mu_0)$. The second step applies Sion's minimax theorem, which permits the interchange of \( \min_{\lambda \in \text{aff}(\Theta)} \) and \( \max_{p \in BP(\mu_0)} \).

The generalized subjective payoff function \( \hat{V}_{\lambda,C}(\mu) \) also reveals the economic logic behind the optimal MDMB, denoted by \( (p^*, x^*) \). Suppose \( \lambda^* \) solves \( \min_{\lambda \in \text{aff}(\Theta)} cav(\hat{V}_{\lambda,C})(\mu_0) \). We can then identify the money-burning amount and the Receiver's best response across two core categories: the \emph{persuasion group} and the \emph{credibility-gaining group}. Given a posterior belief \( \mu \), we say \( \mu \) belongs to the \emph{persuasion group} if \( \omega_{\lambda^*}(\mu)>0 \). At such beliefs, the Sender uses the message for persuasion: the Receiver takes the best response most favorable to the Sender, and no money is burned. Conversely, \( \mu \) falls into the \emph{credibility-gaining group} if \( \omega_{\lambda^*}(\mu)<0 \). These messages are not valuable because they are directly persuasive. They are valuable because they lower deviation payoffs: the Sender burns the maximum amount and induces the Receiver's least favorable best response, which helps make the persuasion group incentive compatible. Based on the above analysis, we present the following proposition that characterizes the structure of the optimal MDMB.

\begin{proposition}\label{prop:opt_MDMB}
    The solution consisting of an MDMB $(p,x)$ and $V$ that is feasible to the program \eqref{eqn:SenderOPT_Canonical} is optimal if and only if there exists $\lambda\in \text{aff}(\Theta)$ such that:
    \begin{enumerate}
        \item[(i)] for any $\mu\in\textbf{supp}\{p\}$ and $\omega_{\lambda}(\mu)>0$, $V(\mu)=\max \mathbb{V}(\mu)$ and $x(\mu)=0$;
        \item[(ii)] for any $\mu\in\textbf{supp}\{p\}$ and $\omega_{\lambda}(\mu)<0$, $V(\mu)=\min \mathbb{V}(\mu)$ and $x(\mu)=C$;
        \item[(iii)] $p\in\ \arg\max_{\tau\in BP(\mu_0)}\int_{\mu}\hat{V}_{\lambda,C}(\mu)d\tau(\mu)$.
    \end{enumerate}
\end{proposition}

The proposition has a saddle-point interpretation. Let 
\[
\mathcal{L}_{C}[\lambda,(p,x,V)]= \int_{\mu}\omega_{\lambda}(\mu)(V(\mu)-x(\mu)) dp(\mu).
\]
Following the discussions below \autoref{thm:SenderOPT_bounded}, we have shown that 
\[
\min_{\lambda\in\text{aff}(\Theta)}\max_{V\in\mathbb{V},x(\mu)\in T,p\in BP(\mu_0)} \mathcal{L}_{C}[\lambda,(p,x,V)] =\max_{V\in\mathbb{V},x(\mu)\in T,p\in BP(\mu_0)} \min_{\lambda\in\text{aff}(\Theta)}\mathcal{L}_{C}[\lambda,(p,x,V)].
\]
Therefore, an optimal solution of program \eqref{eqn:SenderOPT_Canonical}, denoted by $(p^*,x^*,V)$, together with a minimizing multiplier $\lambda$, can be interpreted as a Nash equilibrium of a zero-sum game. Given $\lambda$, conditions (i) and (ii) choose the pointwise best response to the sign of $\omega_{\lambda}(\mu)$, while condition (iii) chooses the best distribution of posteriors. Given a feasible incentive-compatible solution, changing $\lambda\in\text{aff}(\Theta)$ cannot change the value of $\mathcal{L}_{C}$ because \eqref{eqn:IC} makes all type reports yield the same interim payoff. Thus the feasible solution and the associated multiplier form a saddle point, which proves the optimality characterization.

Beyond these two groups, when \( \omega_{\lambda^*}(\mu)=0 \), the message belongs to a third group, which balances interim payoffs to satisfy the incentive compatibility condition \eqref{eqn:IC}. At these beliefs, if the Receiver's action lies between the most preferred and the least preferred best responses, the Sender can reduce the burning cost by slightly shifting the Receiver's mixed strategy toward the Sender's least preferred action. Thus, the following corollary holds: the Sender burns money only when the Receiver takes the least preferred best response.

\begin{corollary}\label{coro:opt_solution}
    There is an optimal solution consisting of an MDMB $(p,x)$ and $V$ to the program \eqref{eqn:SenderOPT_Canonical} such that in equilibrium, $x(\mu)>0$ only if $V(\mu)=\min\mathbb{V}(\mu)$ for all $\mu\in\textbf{supp}\{p\}$.
\end{corollary}

\section{Money Burning Improves Mediated Communication}\label{sec:improvement}

To understand the effect of money burning and the budget $C$, we have to characterize the properties of the credibility-gaining group. In this section, we show that the credibility-gaining group is not only a feature of the optimal design; it is also the source of strict gains from increasing the burning budget.

The intuition comes from the characterization in \autoref{prop:opt_MDMB}. A larger budget matters only at posteriors with negative distorted weight, because these are exactly the posteriors at which the optimal mechanism burns the full amount \(C\). At a differentiability point and holding the optimal posterior distribution fixed, the envelope intuition gives
\[
\frac{\partial \mathcal{V}_C^*(\mu_0)}{\partial C} = -\int_{\mu\in \{\mu: \omega_{\lambda^*}(\mu)<0\}}\omega_{\lambda^*}(\mu) \, dp^*(\mu)
\]
where \( \lambda^* \) denotes the Lagrange multiplier, and \( p^* \) represents the optimal signaling scheme under budget \( C \) in the MDMB problem. Since the integrand is negative exactly on the credibility-gaining group, this expression suggests that increasing \(C\) creates strict value precisely when such posteriors are used.\footnote{This calculation is heuristic because it does not directly apply the envelope theorem. A related rigorous argument is established in the proof of \autoref{thm:credibility_gaining_group} in the appendix.}

To characterize the credibility-gaining group, we impose the following assumption on the Receiver's payoff functions. This assumption holds generically.\footnote{The Lebesgue measure of the set of Receiver payoff functions that do not satisfy this condition is zero in the entire function space.} Whenever this assumption holds, we can show that, generically, the credibility-gaining group is non-empty, implying that money burning improves mediated communication.

\begin{assumption}\label{ass:WeakDominated}
    For any belief $\mu$ and any $a\in RO(\mu)$, there exists $\mu'$ such that $\textbf{supp}\{\mu'\}=\textbf{supp}\{\mu\}$ and $RO(\mu')=\{a\}$.\footnote{This condition is also present in \cite{lipnowski2024perfect}, where they employ it as a generic sufficient criterion for the uniqueness of the Sender's payoff under perfect Bayesian equilibrium in the Bayesian persuasion game.}
\end{assumption}

Let $\mathcal{V}_{CT}^*(\mu_0)$ denote the Sender's maximum payoff under cheap talk equilibrium given the prior $\mu_0$. According to \cite{lipnowski2020cheap}, $\mathcal{V}_{CT}^*(\mu_0)=qcav(\mathbb V)(\mu_0)$ where $qcav(f)$ is the quasi-concave envelope of function $f$. Then, we state the following theorem. When the budget exceeds the range of action payoffs, either additional burning capacity strictly improves the Sender's payoff, or the mediated mechanism with money burning offers no payoff advantage over standard MD and cheap talk.

\begin{theorem}\label{thm:credibility_gaining_group}
    Under \autoref{ass:WeakDominated}, for any $C_2>C_1> \max_{\mu,V(\mu)\in\mathbb{V}(\mu)} V(\mu)-\min_{\mu,V(\mu)\in\mathbb{V}(\mu)} V(\mu)$, either $\mathcal{V}_{C_2}^*(\mu_0)>\mathcal{V}_{C_1}^*(\mu_0)$ or $\mathcal{V}^*_{C_2}(\mu_0)=\mathcal{V}^*_{0}(\mu_0)=\mathcal{V}^*_{CT}(\mu_0)$.
\end{theorem}

\begin{proof}[Proof Overview]
    The proof of \autoref{thm:credibility_gaining_group} formalizes the following idea. If increasing the budget from \(C_1\) to \(C_2\) creates no payoff gain, then an optimal design under \(C_1\) must also be optimal under \(C_2\), and therefore it cannot contain posteriors with negative distorted weight under the larger budget. The remaining posteriors are divided into a persuasion group and a balancing group. The persuasion group must deliver the same Sender payoff across all its posteriors, while the aggregate belief of the balancing group must lie on the boundary of the simplex once the budget exceeds the range of action payoffs. This lowers the dimension of the problem, so the same argument can be applied recursively until the outcome can be represented by a cheap talk equilibrium. The genericity condition in \autoref{ass:WeakDominated} rules out profitable boundary deviations during this recursion.
\end{proof}

\autoref{thm:credibility_gaining_group} characterizes the consequence of having no strict gain from a larger burning budget. In particular, it shows that money burning must expand the equilibrium payoff set and strictly improve the Sender's payoff unless communication via a trusted mediator collapses to that of cheap talk. Building on this characterization, we provide a mild condition under which the credibility-gaining force is guaranteed to be present, thereby ensuring that money burning strictly improves mediated communication.

\begin{proposition}\label{prop:improve_CT_C}
    Under \autoref{ass:WeakDominated}, if $qcav(\mathbb V)(\mu_0)\neq cav(\mathbb V)(\mu_0)$ and there is a sufficiently small $\varepsilon>0$ such that $qcav(\mathbb V)(\mu_0+\varepsilon(\mu-\mu_0))=qcav(\mathbb V)(\mu_0)$ for all $\mu\in \Delta(\Theta)$, it follows that $\mathcal{V}^*_{0}(\mu_0)<\mathcal{V}^*_{C}(\mu_0)$ for any $C>\max_{\mu,V(\mu)\in\mathbb{V}(\mu)} V(\mu)-\min_{\mu,V(\mu)\in\mathbb{V}(\mu)} V(\mu)$.
\end{proposition}

By \autoref{prop:improve_CT_C}, and following a similar argument to Corollary 2 in \cite{lipnowski2020cheap}, we immediately obtain that the value of MDMB serves as a strict upper bound on the Sender's maximum equilibrium payoff achievable under MDMB with finite budgets, for almost all prior beliefs. We summarize this finding in \autoref{coro:generic_better}, which shows that, generically, money burning (without budgetary constraints) strictly improves communication efficiency relative to smaller budgets, including the case of classical mediated communication.

\begin{corollary}\label{coro:generic_better}
    Under \autoref{ass:WeakDominated} and for any $C\in[0,+\infty)$, for almost all prior beliefs $\mu_0\in\Delta(\Theta)$, either $\mathcal{V}^*_{CT}(\mu_0)=\max_{\mu\in\Delta(\Theta)}\max \mathbb V(\mu)$ or $\mathcal{V}^*_{C}(\mu_0)<\mathcal{V}^*(\mu_0)$.
\end{corollary}

\section{The Value of MDMB}\label{sec:value_MDMB}
In this section, we characterize the value of MDMB when the burning budget is unbounded. The finite-budget characterization in \autoref{thm:SenderOPT_bounded} already shows that credibility-gaining posteriors have negative distorted weight. As the budget grows, these negative weights cannot remain bounded away from zero in an optimal multiplier: otherwise the term associated with full burning would dominate the generalized subjective payoff function. Thus, in the limit, the affine multiplier behaves like an ordinary subjective prior in $\Delta(\Theta)$. This observation connects the value of MDMB to robust Bayesian persuasion.

More concretely, if $\lambda_C$ is a minimizing multiplier associated with the optimal MDMB under budget $C$, then the boundedness of the value forces the negative part of $\lambda_C$ to vanish as $C$ becomes large. Hence, although some finite-budget multipliers may satisfy $\lambda_C(\theta)<0$, their negative components converge to zero as the budget increases. When $C=+\infty$, the relevant multipliers can therefore be interpreted as subjective priors.	

We define type $\theta$'s likelihood-adjusted share of the ex-post payoff $\max \mathbb V(\mu)$ at posterior $\mu$ as $\hat{V}_{\theta}(\mu)=\frac{\mu(\theta)}{\mu_{0}(\theta)}\max\mathbb V(\mu)$.\footnote{This follows from the fact that $\mathbb{E}_{\theta\sim\mu_{0}}\{\hat{V}_{\theta}(\mu)\}=\max \mathbb V(\mu)$. The notation $\hat{V}_{\theta}$ is also known as truth-adjusted welfare function introduced by \cite{doval2023persuasion}.} The likelihood ratio $\frac{\mu(\theta)}{\mu_0(\theta)}$ is the weight with which posterior $\mu$ contributes to type $\theta$'s interim payoff. Based on this adjusted ex-post payoff, we introduce the Sender's \emph{subjective payoff function} under the posterior $\mu$ and the subjective prior $\lambda\in\Delta(\Theta)$ as
\begin{equation}\label{eqn:subjective_payoff}
    \hat{V}_{\lambda}(\mu) \triangleq \mathbb{E}_{\theta\sim \lambda} \{\hat{V}_{\theta}(\mu)\} = \sum_{\theta\in\Theta}\lambda(\theta) \frac{\mu(\theta)}{\mu_{0}(\theta)}\max \mathbb V(\mu).\footnote{Note that when $\lambda=\mu_0$ the subjective payoff function under the posterior $\mu$ becomes $\max \mathbb V(\mu)$.}
\end{equation}

Given a canonical MDMB $(\pi,x)$, we define the \emph{interim signaling payoff} of type $\theta$ under signaling scheme $\pi$ as follows,
\begin{equation}\label{eqn:interim_payoff}
    V_{\pi}(\theta)\triangleq \sum_{\mu\in \textbf{supp}\{\pi(\theta)\}} \pi(\mu|\theta)\max \mathbb V(\mu).
\end{equation}

The following proposition characterizes the value of MDMB.
\begin{proposition}\label{thm:min_max} 
$\mathcal{V}^*(\mu_0)=\min_{\lambda\in\Delta(\Theta)}cav(\hat{V}_{\lambda})(\mu_0)=\max_{\pi}\min_{\theta\in \Theta}V_{\pi}(\theta)$.
\end{proposition}

The first equality follows from the limiting version of \autoref{thm:SenderOPT_bounded}: as the negative part of the affine multiplier disappears, the generalized subjective payoff function converges to $\hat V_{\lambda}$ with $\lambda\in\Delta(\Theta)$. The second equality has a direct incentive interpretation. Given a signaling scheme $\pi$, nonnegative money burning implies that any incentive-compatible implementation based on $\pi$ cannot give the Sender a common payoff above $\min_{\theta}V_{\pi}(\theta)$, because every type must be willing to report truthfully and burning can only reduce payoffs. Conversely, the following construction shows that any signaling scheme can be made approximately incentive compatible by adding small-probability credibility-gaining messages and suitable money burning.	

\begin{claim}\label{cl:construct_noise}
Given any signaling scheme $\pi:\Theta\rightarrow \Delta(\Delta(\Theta))$, let $\mu_{\theta}$ denote the degenerate posterior that assigns probability one to type $\theta$. For $\delta\in(0,1)$, construct an MDMB $(\bar{\pi},\bar{x})$ by
$$
\bar{\pi}(\mu|\theta)=(1-\delta)\pi(\mu|\theta)+\delta\mathbf{1}_{\{\mu=\mu_{\theta}\}},
$$
and
$$
\bar{x}(\mu)=\left\{
\begin{matrix}
    \frac{V_{\bar{\pi}}(\theta)-\min_{\theta'\in\Theta}V_{\bar{\pi}}(\theta')}{\bar{\pi}(\mu_{\theta}|\theta)} & \mu=\mu_{\theta}\text{ for some }\theta\in\Theta,\\
    0 & \text{otherwise.}
\end{matrix}
\right.
$$
The mechanism $(\bar{\pi},\bar{x})$ is incentive-compatible for $\delta\in(0,1)$, and the Sender's payoff under this mechanism converges to $\min_{\theta\in\Theta}V_{\pi}(\theta)$ as $\delta\rightarrow 0^+$.
\end{claim}

The construction retains the original signaling scheme with probability close to one and adds type-revealing messages with small probability. These additional messages are credibility-gaining messages: they need not be attractive on their own, but they allow the mechanism to equalize interim payoffs across types. Although this construction may not be optimal for any finite $C$ (as shown in \autoref{prop:opt_MDMB} since the money-burning amounts differ across messages), as $C \to \infty$, the messages outside the persuasion group play an increasingly negligible role in the optimal design. Therefore, the construction in \autoref{cl:construct_noise} provides a good approximation, and the Sender's payoff under this construction converges to the value of MDMB as $\delta \to 0^+$ and $C \to +\infty$.

This section also highlights that when $C$ is sufficiently large, the persuasion group consists of posteriors supported by the signaling scheme that maximizes the Sender's minimum interim payoff. Moreover, the Sender's payoff approaches $\mathcal{V}^*(\mu_0)$, as characterized in \autoref{thm:min_max}, as $C \to \infty$.

\subsection{Money Burning Secures Commitment}
We demonstrate that the value of MDMB can be interpreted as the value of two robust Bayesian persuasion problems. In both interpretations, the Sender has full commitment power but evaluates signaling schemes through a worst-case criterion. Therefore, compared to classical MD, MDMB implements part of the commitment value of Bayesian persuasion through the money-burning instrument. It also provides a theoretical micro-foundation for max-min persuasion objectives.

The max-min representation in \autoref{thm:min_max} indicates that the value of MDMB is the same as the value of a Sender who has full commitment power and chooses a signaling scheme to maximize his minimum interim payoff across types. This problem is cautious Bayesian persuasion: the Sender has full commitment power but acts robustly to the type realization \cite{doval2021information,doval2023persuasion}. The following corollary constitutes our first implication of \autoref{thm:min_max}.
\begin{corollary}\label{coro:cautious_BP}
    The value of MDMB equals the payoff of the Sender with full commitment power but who is cautious.
\end{corollary}

The minimization over subjective priors in \autoref{thm:min_max} relates the value of MDMB to robust Bayesian persuasion with heterogeneous beliefs. We call a subjective distribution $\lambda\in\Delta(\Theta)$ the \emph{worst Sender's subjective prior} if it minimizes $cav(\hat{V}_{\lambda})(\mu_0)$, and we refer to this minimized value as \emph{the worst Sender's subjective expected payoff}. Based on the model introduced by \cite{alonso2016bayesian}, in which the Sender's and the Receiver's subjective priors are heterogeneous, \autoref{thm:min_max} shows that the value of MDMB coincides with the Sender's payoff under the worst subjective prior. Hence, the following corollary is our second implication.

\begin{corollary}\label{coro:heteBP}
The value of MDMB $\mathcal{V}^*(\mu_0)$ equals the payoff of the Sender under Bayesian persuasion with heterogeneous priors, in which the Sender holds the worst subjective prior and the Receiver has prior $\mu_0$.
\end{corollary}

\subsection{More on the Persuasion Group}\label{sec:construction}
In \autoref{sec:value_MDMB}, we have examined the properties of the persuasion group under a sufficiently large budget. Note that \autoref{thm:min_max} also provides a minimax characterization, where the limiting Lagrange multiplier is a nonnegative subjective prior. Accordingly, in this section, we further characterize the persuasion group under a large budget. According to \autoref{thm:min_max}, the key object is a signaling scheme that maximizes the minimum interim payoff, which we refer to as \emph{the optimal signaling scheme}. We can compute the optimal signaling scheme from the saddle-point relation between this signaling scheme and the worst subjective prior.

The worst subjective belief and the optimal signaling scheme can be modeled as a Nash equilibrium in a zero-sum game between the Sender and Nature. In this game, the Sender chooses a signaling scheme $p \in BP(\mu_0)$, while Nature selects the Sender's subjective prior $\lambda \in \Delta(\Theta)$. The Sender seeks to maximize the following payoff function, whereas Nature aims to minimize it.
\begin{equation}
    \mathcal{L}(\lambda,p)\triangleq\int_{\mu} \hat{V}_{\lambda}(\mu)dp(\mu).
\end{equation}
This function represents the Sender's subjective expected payoff under the signaling scheme $p$ and the prior belief $\lambda$. The following propositions state the indifference condition that characterizes the saddle point of the zero-sum game.
\begin{proposition}\label{prop:opt_lambda}
A subjective prior $\lambda^*\in\Delta(\Theta)$ is the worst Sender's subjective prior if and only if there exists $p^*\in BP(\mu_0)$ such that $\mathcal{L}(\lambda^*,p^*)=cav(\hat{V}_{\lambda^*})(\mu_0)$, and for any $\theta\in\textbf{supp}(\lambda^*)$, $\mathcal{L}(\lambda^*,p^*)=\mathcal{L}(\mu_{\theta},p^*)$, and for any $\theta\not\in\textbf{supp}(\lambda^*)$, $\mathcal{L}(\lambda^*,p^*)\leq \mathcal{L}(\mu_{\theta},p^*)$.\footnote{$\mu_{\theta}$ is the distribution in $\Delta(\Theta)$ with a singleton support $\{\theta\}$.}
\end{proposition}

\begin{proposition}\label{prop:OPTp}
A signaling scheme $p^*\in BP(\mu_0)$ is optimal if and only if there exists $\lambda^*\in \Delta(\Theta)$ such that $\mathcal{L}(\lambda^*,p^*)=cav(\hat{V}_{\lambda^*})(\mu_0)$, and for any $\theta\in\textbf{supp}(\lambda^*)$, $\mathcal{L}(\lambda^*,p^*)=\mathcal{L}(\mu_{\theta},p^*)$, and for any $\theta\not\in\textbf{supp}(\lambda^*)$, $\mathcal{L}(\lambda^*,p^*)\leq \mathcal{L}(\mu_{\theta},p^*)$.
\end{proposition}

These two propositions jointly characterize the worst subjective prior and the optimal signaling scheme. At a saddle point, Nature puts positive probability only on types whose interim signaling payoffs are minimal; types outside Nature's support have weakly higher interim payoffs. Thus the persuasion group under a large budget is pinned down by the signaling scheme that survives this max-min indifference condition.

We can apply the characterization of the worst subjective prior to narrow the possible range of the set $\arg\min_{\lambda\in\Delta(\Theta)}cav(\hat{V}_{\lambda})(\mu_0)$ when there are only two possible types of the Sender in the appendix.

\subsection{The Salesman Example}

At the end of this section, we use our previous results to compare the MDMBs with different budgets, Bayesian persuasion, mediated communication, and cheap talk using \autoref{example:illustrative}.

We begin by characterizing $\mathcal{V}^*_{C}(\mu_0)$. By \autoref{thm:SenderOPT_bounded}, we need to consider four lines for a given parameter $\lambda\in\mathbb{R}$, namely
$
l_1(\mu)=0,l_2(\mu)=-C(\frac{\lambda\mu}{\mu_0}+\frac{(1-\lambda)(1-\mu)}{1-\mu_0}), l_3(\mu)=\frac{\lambda\mu}{\mu_0}+\frac{(1-\lambda)(1-\mu)}{1-\mu_0}
$ and $l_4(\mu)=(1-C)(\frac{\lambda\mu}{\mu_0}+\frac{(1-\lambda)(1-\mu)}{1-\mu_0})$. Correspondingly, we have that for $\mu\in[0,1]$,
$$
\hat{V}_{\lambda,C}(\mu)=\left\{
\begin{matrix}
\max\{l_1(\mu),l_2(\mu)\} & \mu<\frac{1}{2}; \\
\max\{l_2(\mu),l_3(\mu)\} & \mu=\frac{1}{2}; \\
\max\{l_3(\mu),l_4(\mu)\} & \mu>\frac{1}{2}.
\end{matrix}
\right.
$$
Since $\hat{V}_{\lambda,C}$ is convex and upper semi-continuous on $\mu\in[0,\frac{1}{2})$ and $\mu\in(\frac{1}{2},1]$, to compute $cav(\hat{V}_{\lambda,C})(\mu_0)$ we only need to evaluate $\hat{V}_{\lambda,C}(0)=\max\{0,-C\frac{1-\lambda}{1-\mu_0}\}$, $\hat{V}_{\lambda,C}(\frac{1}{2})=\max\{\frac{-C}{2}(\frac{\lambda}{\mu_0}+\frac{1-\lambda}{1-\mu_0}),\frac{1}{2}(\frac{\lambda}{\mu_0}+\frac{1-\lambda}{1-\mu_0})\}$ and $\hat{V}_{\lambda,C}(1)=\max\{\frac{\lambda}{\mu_0},(1-C)\frac{\lambda}{\mu_0}\}$. Assuming $\mu_0<\frac{1}{2}$, we can only partition $\mu_0$ into $0,\frac{1}{2}$ or $0,1$. Since $\hat{V}_{\lambda,C}(0),\hat{V}_{\lambda,C}(\frac{1}{2}),\hat{V}_{\lambda,C}(1)$ are all decreasing in $\lambda$ for $\lambda\geq 0$, to find the minimum concavification value, we only need to consider the case of $\lambda\leq 0$. We then divide this case into two subcases: $\lambda\in[-\frac{\mu_0}{1-2\mu_0},0]$ and $\lambda\in(-\infty,-\frac{\mu_0}{1-2\mu_0})$. By considering two subcases, we solve the Lagrange multiplier $\lambda^*=\frac{-\mu_0}{(1-\mu_0)C-\mu_0}$ for $C>1$. Thus, we can solve for the result and obtain that for $\mu_0< \frac{1}{2}$,
$$
\mathcal{V}^*_{C}(\mu_0)=
\left\{
\begin{matrix}
    0 & C\leq 1; \\
   \frac{(C-1)\mu_0}{C(1-\mu_0)-\mu_0} & C>1.
\end{matrix}
\right.
$$

We thus show the optimal Sender's payoffs of different communication protocols in \autoref{fig:Vcost}. The red line is the concave envelope of $\max \mathbb{V}(\mu)$, which is the result of Bayesian persuasion. The black line is the result of the MDMB with bound $C=+\infty$, i.e. $\mathcal{V}^*(\mu_0)$. The blue line is the result of $\mathcal{V}^*_{2}(\mu_0)$. Finally, we can see that regardless of what we use among the MDMB with bounded budget $C \leq 1$, cheap talk, or classical mediated communication, we can only get the results as the green line, which cannot benefit from those protocols. 

\begin{figure}[!htb]
    \centering
    \begin{tikzpicture}[scale=5]
    \draw[->](-0.1,0)--(1.2,0)node[left,below]{$\mu_0$};
    \draw[->](0,-0.1)--(0,0.8)node[right]{$V$};
    \draw[domain=0:0.5, thick]
            plot(\x,{0.6*\x/(1-\x)});
    \draw[thick] (0.5,0.6)--(1,0.6);

    \draw[color=red,thick] (0,0)--(0.5,0.6)--(1,0.6);

    \draw[domain=0:0.5, thick, color=blue]
            plot(\x,{0.6*\x/((2-3*\x)}); %C=2

    %\draw[domain=0:0.5, thick, color=green]
         %   plot(\x,{0.6*0.3*\x/((1.3-2.3*\x)}); %C=1.3

    \draw[thick, color=green] (0,0)--(0.5,0); %C<=1 or cheap talk

    \draw(0,0)node[below left]{0};
    \draw[dashed]  (1,0.6)--(1,0)node[below]{1};
    \draw[dashed]  (0.5,0.6)--(0.5,0)node[below]{0.5};
    \draw[dashed]  (0.5,0.6)--(0,0.6)node[left]{1};

    \end{tikzpicture}
    \caption{Comparison of Different Protocols}
    \label{fig:Vcost}
\end{figure}

\section{Discussions}\label{sec:discuss}
\subsection{The Value of Commitment in Web 3 Communities} \label{sec:compare}
The previous sections identify MDMB as an intermediate commitment technology: stronger than cheap talk because a mediator can enforce report-contingent consequences, but weaker than Bayesian persuasion because the Sender still privately observes his type and chooses his report. This distinction is useful for Web 3.0 environments, where smart contracts and transparent algorithms can enforce communication rules and observable costs without giving the Sender full commitment power. Let $\mathcal{V}^*_{BP}(\mu_0)$ represent the value of Bayesian persuasion; that is, $\mathcal{V}^*_{BP}(\mu_0)=cav(\mathbb V)(\mu_0)$. This section compares $\mathcal{V}^*_{BP}(\mu_0)$ with $\mathcal{V}^*(\mu_0)$, which captures the value achievable through MDMB-like commitment.

In conventional environments, communication without commitment is often modeled as cheap talk. Consequently, $\mathcal{V}^*_{BP}(\mu_0)-\mathcal{V}^*_{CT}(\mu_0)$ quantifies the conventional value of commitment. In Web 3.0 communities, as discussed by \cite{drakopoulos2022blockchain}, smart contracts can serve as transparent mediators: they can implement pre-specified algorithms, make outputs observable, and impose report-contingent costs through gas fees, token transfers, or subsidies. These technologies move the relevant benchmark away from pure cheap talk and toward MDMB. Hence, we define $\mathcal{V}^*_{BP}(\mu_0)-\mathcal{V}^*(\mu_0)$ as the \emph{refined value of commitment} in Web 3.0 communities. This gap is the remaining cost of not having full commitment after algorithmic mediation and enforceable burning are available.

Our first result establishes the condition under which the refined value of commitment vanishes.
\begin{proposition}\label{prop:allequal}
If $\mathcal{V}^*(\mu_0)=\mathcal{V}^*_{BP}(\mu_0)$ then $\mathcal{V}^*_{CT}(\mu_0)=\mathcal{V}^*_{0}(\mu_0)=\mathcal{V}^*(\mu_0)= \mathcal{V}^*_{BP}(\mu_0)$.
\end{proposition}
Intuitively, if MDMB already attains the Bayesian persuasion value, then algorithmic mediation and enforceable burning close the entire commitment gap. In that case, the comparison result of \cite{corrao2023communication} implies that standard mediated communication and cheap talk must also attain the Bayesian persuasion value. By \autoref{prop:allequal}, if $\mathcal{V}_{CT}^*(\mu_0)<\mathcal{V}^*_{BP}(\mu_0)$, then $\mathcal{V}^*(\mu_0)<\mathcal{V}^*_{BP}(\mu_0)$. Consequently, we obtain the following corollary, which compares the conventional value of commitment with the refined value of commitment.
\begin{corollary}\label{coro:VneqBP}
There is a positive value of commitment, i.e. $\mathcal{V}^*_{CT}(\mu_0)<\mathcal{V}^*_{BP}(\mu_0)$, if and only if there is a positive refined value of commitment in the Web 3.0 community, i.e. $\mathcal{V}^*(\mu_0)<\mathcal{V}^*_{BP}(\mu_0)$. 
\end{corollary}

The equivalence in \autoref{coro:VneqBP} does not imply that MDMB always narrows the commitment gap. In \autoref{example:CT=BD=BM}, we show that the relation $\mathcal{V}^*_{CT}(\mu_0)=\mathcal{V}^*_{0}(\mu_0)=\mathcal{V}^*(\mu_0)< \mathcal{V}^*_{BP}(\mu_0)$ can hold even under \autoref{ass:WeakDominated}. This is the case in which the refined value of commitment is positive and equal to the conventional value of commitment.

\autoref{coro:VneqBP} has an important implication: if commitment is valuable in conventional communication environments, then full commitment remains valuable even in Web 3.0 communities with algorithmic mediation and enforceable burning. At the same time, whenever MDMB improves on the cheap-talk benchmark, the refined value of commitment is smaller than the conventional value. Thus algorithmic mediation and enforceable burning can mitigate, but generally do not eliminate, the loss from the absence of full commitment. In addition, according to Corollary 2 in \cite{lipnowski2020cheap}, the refined value of commitment is strictly positive under almost all prior beliefs as long as the value of commitment is strictly positive. 

\subsection{Comparison with Money Burning in Cheap Talk}\label{sec:ct_money_burning}
Numerous studies have explored the role of money burning in cheap talk equilibria \cite{austen2000cheap, kartik2007note,karamychev2017optimal}, where money burning is a voluntary costly signal chosen by the Sender. These studies show that money burning can enhance the precision of a cheap talk equilibrium, but it makes cheap talk influential\footnote{An equilibrium is influential if there are at least two of the Receiver's actions taken along the equilibrium path with the same amount of money burning.} only if some cheap talk equilibrium without money burning is already influential. Another limitation is that, although money burning can expand the equilibrium set of cheap talk in some cases, the Sender may dissipate all informational gains through the burning cost.

Under transparent motives, this limitation is stark: the Sender cannot use voluntary money burning to improve his payoff in a cheap talk equilibrium. In the binary-type example, \autoref{example:illustrative}, when $\mu_0 < 0.5$, the Sender cannot achieve a higher payoff than the no-communication payoff under cheap talk, cheap talk with money burning, or classical mediated communication. Money burning can still expand the cheap-talk equilibrium set; for example, burning a payoff of $1$ at posterior $\mu=1$ can induce a separating equilibrium. But it does not improve the Sender's payoff. In MDMB, by contrast, burning is an enforceable consequence of a mediated report. It can therefore serve as an incentive instrument: credibility-gaining messages discipline deviations and make more persuasive messages incentive compatible. This distinction explains why money burning can enhance the precision of cheap talk without improving the Sender's credibility, as in Theorems 1 and 2 of \cite{kartik2007note}, whereas \autoref{thm:credibility_gaining_group} shows that money burning can expand the equilibrium payoff set in mediated communication.

Beyond the assumption of transparent motives, a study on the implementation problem \cite{wu2023eliciting} demonstrates a significant difference in the implementability conditions between implementing a cheap talk equilibrium with money burning and a mediated communication equilibrium with money burning. In addition, in \autoref{example:Beyond_Trans_Mot}, we show that money burning cannot always make mediated communication credible, implying that money burning cannot enhance mediated communication and the necessity of the transparent motives assumption.

\section{Concluding Remarks}\label{sec:conlu}
This paper asks when a purely wasteful action can improve strategic communication. The answer is not that money burning is valuable by itself. Without commitment, burning is only a voluntary costly signal and generally cannot improve on cheap talk. With full commitment, burning is redundant because the Sender can directly commit to an information structure. Money burning becomes useful at an intermediate level of commitment, where a mediator can enforce report-contingent consequences but the Sender still privately observes his type and chooses his report.

We formalize this intermediate environment through mediated communication with money burning (MDMB). By developing a revelation principle tailored to observable money burning, we reduce the problem to a belief-based program with incentive-compatibility, obedience, Bayes-plausibility, and budget constraints. The optimal mechanism separates posterior messages into a persuasion group and a credibility-gaining group. Persuasion-group messages induce the Receiver's best responses most favorable to the Sender and burn no money; credibility-gaining messages burn resources and induce less favorable actions, but they discipline deviations and make the persuasive messages credible.

The main result shows that this force is generically strict. Once the burning budget exceeds the range of action payoffs, increasing the budget strictly improves the Sender's payoff unless the value of MDMB collapses to the cheap-talk benchmark. Thus money burning does not merely refine communication equilibria. In a mediated environment, it can expand the Sender's equilibrium payoff set and strictly improve mediated communication whenever commitment has value for almost all priors.

In practice, this framework clarifies the role of algorithmic mediation in Web 3.0 communities. Smart contracts and transparent algorithms can implement MDMB-like communication by enforcing pre-specified rules and report-contingent costs. They can therefore mitigate the loss from the absence of full commitment, but they generally do not eliminate it. The remaining gap between Bayesian persuasion and MDMB is the refined value of commitment.

These conclusions rely on the transparent-motives assumption. In environments with state-dependent Sender preferences, money burning may interact with incentives in different ways. Deriving sharper quantitative bounds on the payoff improvement from money burning, and extending the analysis beyond transparent motives, remain natural directions for future work.

\bibliography{used_device}

@article{kamenica2011bayesian,
  title={Bayesian persuasion},
  author={Kamenica, Emir and Gentzkow, Matthew},
  journal={American Economic Review},
  volume={101},
  number={6},
  pages={2590--2615},
  year={2011}
}

@article{lipnowski2020cheap,
  title={Cheap talk with transparent motives},
  author={Lipnowski, Elliot and Ravid, Doron},
  journal={Econometrica},
  volume={88},
  number={4},
  pages={1631--1660},
  year={2020},
  publisher={Wiley Online Library}
}

@article{salamanca2021value,
  title={The value of mediated communication},
  author={Salamanca, Andr{\'e}s},
  journal={Journal of Economic Theory},
  volume={192},
  pages={105191},
  year={2021},
  publisher={Elsevier}
}

@article{crawford1982strategic,
  title={Strategic information transmission},
  author={Crawford, Vincent P and Sobel, Joel},
  journal={Econometrica: Journal of the Econometric Society},
  pages={1431--1451},
  year={1982},
  publisher={JSTOR}
}

@article{forges1986approach,
  title={An approach to communication equilibria},
  author={Forges, Francoise},
  journal={Econometrica: Journal of the Econometric Society},
  pages={1375--1385},
  year={1986},
  publisher={JSTOR}
}

@article{austen2000cheap,
  title={Cheap talk and burned money},
  author={Austen-Smith, David and Banks, Jeffrey S},
  journal={Journal of Economic Theory},
  volume={91},
  number={1},
  pages={1--16},
  year={2000},
  publisher={Elsevier}
}

@article{lipnowski2022persuasion,
  title={Persuasion via weak institutions},
  author={Lipnowski, Elliot and Ravid, Doron and Shishkin, Denis},
  journal={Journal of Political Economy},
  volume={130},
  number={10},
  pages={2705--2730},
  year={2022},
  publisher={The University of Chicago Press Chicago, IL}
}

@article{min2021bayesian,
  title={Bayesian persuasion under partial commitment},
  author={Min, Daehong},
  journal={Economic Theory},
  volume={72},
  number={3},
  pages={743--764},
  year={2021},
  publisher={Springer}
}

@article{lin2022credible,
  title={Credible persuasion},
  author={Lin, Xiao and Liu, Ce},
  journal={Journal of Political Economy},
  volume={132},
  number={7},
  pages={2228-2273},
  year={2024},
  publisher={The University of Chicago Press Chicago, IL}
}

@article{myerson1982optimal,
  title={Optimal coordination mechanisms in generalized principal--agent problems},
  author={Myerson, Roger B},
  journal={Journal of Mathematical Economics},
  volume={10},
  number={1},
  pages={67--81},
  year={1982},
  publisher={Elsevier}
}

@inproceedings{drakopoulos2022blockchain,
  title={Blockchain Mediated Persuasion},
  author={Drakopoulos, Kimon and Lo, Irene and Mulvany, Justin},
  booktitle={Proceedings of the 24th ACM Conference on Economics and Computation},
  year={2023}
}

@article{bergemann2019information,
  title={Information design: A unified perspective},
  author={Bergemann, Dirk and Morris, Stephen},
  journal={Journal of Economic Literature},
  volume={57},
  number={1},
  pages={44--95},
  year={2019}
}

@article{ivanov2014beneficial,
  title={Beneficial mediated communication in cheap talk},
  author={Ivanov, Maxim},
  journal={Journal of Mathematical Economics},
  volume={55},
  pages={129--135},
  year={2014},
  publisher={Elsevier}
}

@article{goltsman2009mediation,
  title={Mediation, arbitration and negotiation},
  author={Goltsman, Maria and H{\"o}rner, Johannes and Pavlov, Gregory and Squintani, Francesco},
  journal={Journal of Economic Theory},
  volume={144},
  number={4},
  pages={1397--1420},
  year={2009},
  publisher={Elsevier}
}

@article{chakraborty2010persuasion,
  title={Persuasion by cheap talk},
  author={Chakraborty, Archishman and Harbaugh, Rick},
  journal={American Economic Review},
  volume={100},
  number={5},
  pages={2361--2382},
  year={2010},
  publisher={American Economic Association}
}

@article{kartik2007note,
  title={A note on cheap talk and burned money},
  author={Kartik, Navin},
  journal={Journal of Economic Theory},
  volume={136},
  number={1},
  pages={749--758},
  year={2007},
  publisher={Elsevier}
}

@article{karamychev2017optimal,
  title={Optimal signaling with cheap talk and money burning},
  author={Karamychev, Vladimir and Visser, Bauke},
  journal={International Journal of Game Theory},
  volume={46},
  number={3},
  pages={813--850},
  year={2017},
  publisher={Springer}
}

@article{sadakane2023multistage,
  title={Multistage information transmission with voluntary monetary transfers},
  author={Sadakane, Hitoshi},
  journal={Theoretical Economics},
  volume={18},
  number={1},
  pages={267--301},
  year={2023}
}

@article{kolotilin2021relational,
  title={Relational communication},
  author={Kolotilin, Anton and Li, Hongyi},
  journal={Theoretical Economics},
  volume={16},
  number={4},
  pages={1391--1430},
  year={2021},
  publisher={Wiley Online Library}
}

@article{li2017discriminatory,
  title={Discriminatory information disclosure},
  author={Li, Hao and Shi, Xianwen},
  journal={American Economic Review},
  volume={107},
  number={11},
  pages={3363--3385},
  year={2017},
  publisher={American Economic Association 2014 Broadway, Suite 305, Nashville, TN 37203}
}

@article{bergemann2018design,
  title={The design and price of information},
  author={Bergemann, Dirk and Bonatti, Alessandro and Smolin, Alex},
  journal={American Economic Review},
  volume={108},
  number={1},
  pages={1--48},
  year={2018},
  publisher={American Economic Association 2014 Broadway, Suite 305, Nashville, TN 37203}
}

@article{frechette2022rules,
  title={Rules and commitment in communication: An experimental analysis},
  author={Fr{\'e}chette, Guillaume R and Lizzeri, Alessandro and Perego, Jacopo},
  journal={Econometrica},
  volume={90},
  number={5},
  pages={2283--2318},
  year={2022},
  publisher={Wiley Online Library}
}

@article{corrao2023communication,
  title={The Bounds of Mediated Communication},
  author={Corrao, Roberto and Dai, Yifan},
  journal={Working Paper},
  year={2023}
}

@article{alonso2016bayesian,
  title={Bayesian persuasion with heterogeneous priors},
  author={Alonso, Ricardo and C{\^a}mara, Odilon},
  journal={Journal of Economic Theory},
  volume={165},
  pages={672--706},
  year={2016},
  publisher={Elsevier}
}

@article{perez2022test,
  title={Test design under falsification},
  author={Perez-Richet, Eduardo and Skreta, Vasiliki},
  journal={Econometrica},
  volume={90},
  number={3},
  pages={1109--1142},
  year={2022},
  publisher={Wiley Online Library}
}

@article{wu2023eliciting,
  title={Implementing Randomized Allocation Rules with Outcome-contingent Transfers},
  author={Liu, Yi and Wu, Fan},
  journal={Journal of Economic Theory},
  volume = {220},
    pages = {105878},
    year = {2024},
    publisher={Elsevier}
}

@article{doval2023persuasion,
  title={Persuasion and Welfare},
  author={Doval, Laura and Smolin, Alex},
  journal={Journal of Political Economy},
  volume={132},
  number={7},
  pages={2451-2487},
  year={2024},
  publisher={The University of Chicago Press Chicago, IL}
}

@article{doval2021information,
  title={Information payoffs: An interim perspective},
  author={Doval, Laura and Smolin, Alex},
  journal={arXiv preprint arXiv:2109.03061},
  year={2021}
}

@article{doval2022mechanism,
author = {Doval, Laura and Skreta, Vasiliki},
title = {Mechanism Design With Limited Commitment},
journal = {Econometrica},
volume = {90},
number = {4},
pages = {1463-1500},
year = {2022}
}

@article{corrao2023mediation,
  title={Mediation Markets: The Case of Soft Information},
  author={Corrao, Roberto},
  year={2023},
  journal={Working Paper}
}

@article{bester2001contracting,
  title={Contracting with imperfect commitment and the revelation principle: the single agent case},
  author={Bester, Helmut and Strausz, Roland},
  journal={Econometrica},
  volume={69},
  number={4},
  pages={1077--1098},
  year={2001},
  publisher={Wiley Online Library}
}

@article{lipnowski2024perfect,
  title={Perfect Bayesian Persuasion},
  author={Lipnowski, Elliot and Ravid, Doron and Shishkin, Denis},
  journal={Journal of Political Economy: Microeconomics},
  year={Forthcoming}
}

@article{shaker2021online,
  title={Online rating system development using blockchain-based distributed ledger technology},
  author={Shaker, Monir and Shams Aliee, Fereidoon and Fotohi, Reza},
  journal={Wireless Networks},
  volume={27},
  number={3},
  pages={1715--1737},
  year={2021},
  publisher={Springer}
}

@article{gersbach2004money,
  title={The money-burning refinement: with an application to a political signalling game},
  author={Gersbach, Hans},
  journal={International Journal of Game Theory},
  volume={33},
  pages={67--87},
  year={2004},
  publisher={Springer}
}

@article{koessler2026belief,
  title={A belief-based approach to signaling},
  author={Koessler, Fr{\'e}d{\'e}ric and Laclau, Marie and Tomala, Tristan},
  year={2026},
  journal={Working Paper}
}
\bibliographystyle{plainnat}

\newpage

\appendix

\section*{\centering{Appendix}}

\section{Omitted Examples}\label{sec:appendix_examples}

\subsection{The Necessity of \autoref{ass:WeakDominated}}

In this section, we provide an example to show how the results in \autoref{sec:improvement} fail if \autoref{ass:WeakDominated} does not hold. 

\begin{example}\label{example:BM=BD>CT}
We present an abstract setting in this example, where we only specify the belief-value function and ensure the existence of the basic settings of $A, u, v, \Theta$ by imposing the upper-semi continuity of the belief-value function.

We assume that there are three distinct types $\theta_1,\theta_2$ and $\theta_3$. The maximum of belief-value correspondence is
$$
V(\mu)=\left\{
\begin{matrix}
    \frac{7}{3} & \mu(\theta_1)=1\\
    2 & \mu(\theta_1)=0, \mu(\theta_2)\in [0,\frac{1}{2}) \\
    3 & \mu(\theta_1)=0, \mu(\theta_2)\in [\frac{1}{2},\frac{3}{4}] \\
    1 & \mu(\theta_1)=0, \mu(\theta_2)\in (\frac{3}{4},1] \\
    1 & \rm{otherwise}
\end{matrix}
\right..
$$
\end{example}
By restricting the support to $\{\theta_2,\theta_3\}$, the value function of \autoref{example:BM=BD>CT} coincides with Example 3 or Figure 7 in \cite{salamanca2021value}. For $\mu_0=(\frac{1}{2},\frac{1}{6},\frac{1}{3})$, since $\mathcal{V}^*_{0}((0,\frac{1}{3},\frac{2}{3}))=\frac{7}{3}$ as shown by \cite{salamanca2021value}, splitting $\mu_0$ into $(1,0,0)$ and $(0,\frac{1}{3},\frac{2}{3})$ yields $\mathcal{V}_{MD}^*(\mu_0)=\frac{7}{3}$. Furthermore, the interim payoff of $\theta_1$ cannot exceed $\frac{7}{3}$, implying that $\mathcal{V}^*(\mu_0)\leq \frac{7}{3}$. Hence, we obtain $\mathcal{V}^*(\mu_0)=\mathcal{V}_{MD}^*(\mu_0)=\frac{7}{3}$. However, to find a cheap talk equilibrium with $\frac{7}{3}$ as the Sender's payoff, we need to split $\mu_0$ into $(1,0,0)$ and $(0,\frac{1}{3},\frac{2}{3})$ and keep $(1,0,0)$ unchanged. Since $\mathcal{V}_{CT}^*((0,\frac{1}{3},\frac{2}{3}))=2<\frac{7}{3}=V((1,0,0))$, no cheap talk equilibrium achieves $\frac{7}{3}$ for the Sender, and thus $\mathcal{V}_{CT}^*(\mu_0)<\mathcal{V}_{MD}^*(\mu_0)=\mathcal{V}^*(\mu_0)$ for $\mu_0=(\frac{1}{2},\frac{1}{6},\frac{1}{3})$.

\subsection{The Necessity of Transparent-motives Assumption}
In this section, we provide an example to show that without the transparent-motives assumption, money burning may not improve mediated communication at all. 
\begin{example}\label{example:Beyond_Trans_Mot}
Consider a buyer (Sender) and a seller (Receiver) in a market. The buyer's valuation of the seller's product, which is the buyer's private information, is drawn from a set \(\Theta = \{1, 2\}\). The probability that the valuation \(v\) is 2 is given by \(\mu_0\).

The seller can set a price \(p\) chosen from the set \(A = \{1, 2\}\). When the seller sets a price \(p\) and the buyer’s valuation is \(v\), the payoffs are defined as follows:

\begin{itemize}
    \item \textbf{Buyer's payoff}: \(v(p, v) = \max\{0, v - p\}\), representing the buyer's surplus.
    \item \textbf{Seller's payoff}: \(u(p, v) = \mathbb{I}(v \geq p) \cdot p\), where \(\mathbb{I}(v \geq p)\) is an indicator function equal to 1 if \(v \geq p\) and 0 otherwise, representing the seller's revenue.
\end{itemize}
\end{example}

In this example, we show that for any \(\mu_0 > 0.5\), there is no MDMB that can induce an outcome where the seller sets a price of 1. Therefore, MDMB cannot improve the buyer's total payoff; however, BP can. Further, we show that even if the money burning is private, which is a stronger setting, the money burning still cannot improve mediated communication. In this setting, we can merge the posteriors that induce the same action of the seller. It is without loss of generality to assume that there are two possible posteriors \( x_1 < x_2 \) of the seller induced by an MDMB. If the MDMB can improve the Sender's payoff, then we must have \( x_1 \leq 0.5 \leq x_2 \). 

If the unconditional probability of the posterior $x_1$ is $p$, then we must have $px_1+(1-p)x_2=\mu_0$. Suppose the expected amount of money burned by a valuation-1 buyer is $T(1)$, and the expected amount of money burned by a valuation-2 buyer is $T(2)$. Then, the incentive-compatible constraints can be written as 
$$
T(2)\geq T(1)
$$
and 
$$
\frac{px_1}{px_1+(1-p)x_2} -T(2)\geq \frac{p(1-x_1)}{p(1-x_1)+(1-p)(1-x_2)} -T(1).
$$
Thus, we can deduce that  
$$
\frac{px_1}{px_1+(1-p)x_2}\geq \frac{p(1-x_1)}{p(1-x_1)+(1-p)(1-x_2)},
$$
which is equivalent to $x_1\geq x_2$. A contradiction!

\subsection{No Value of MDMB}
In this section, we provide an example showing that the MDMB may not improve the communication efficiency even under \autoref{ass:WeakDominated}. Thus, the phrases, ``almost all" or ``generically", in \autoref{sec:improvement} are necessary. 

\begin{example}\label{example:CT=BD=BM}
The Receiver has three possible actions $a_1,a_2,a_3$ and the Sender has two possible types $H,L$. The prior belief assigns probability $\mu_0$ to the Sender's type being $H$. The Sender's values for the actions are $v(a_1)=0,v(a_2)=\frac{1}{4},v(a_3)=1$. We summarize the Receiver's payoffs in \autoref{tab:Rpayoff}.
\begin{table}[htbp]
    \centering
    \begin{tabular}{|c|c|c|}
    \hline
    $u(a,\theta)$& H & L \\ \hline
    $a_1$        & -4 & 1 \\\hline
    $a_2$        & 0 & 0  \\ \hline
    $a_3$        & 1 & -2 \\ \hline
\end{tabular}
    \caption{Receiver's payoff matrix.}
    \label{tab:Rpayoff}
\end{table}

In this example, the belief-value correspondence is
$$
\mathbb{V}(\mu)=\left\{
\begin{matrix}
    1 & \mu\in(\frac{2}{3},1] \\
    [\frac{1}{4},1] & \mu = \frac{2}{3} \\
    \frac{1}{4} & \mu\in(\frac{1}{5},\frac{2}{3}) \\
    [0,\frac{1}{4}] & \mu=1/5\\
    0 & \mu<\frac{1}{5}
\end{matrix}
\right..
$$
\end{example}

\begin{figure}[!htb]
    \centering
    \begin{tikzpicture}[scale=5]
    \draw[->](-0.1,0)--(1.2,0)node[left,below]{$\mu_0$};
    \draw[->](0,-0.1)--(0,0.8)node[right]{$V$};
    %\draw[domain=0:2/3, thick]
        %    plot(\x,{0.6*\x/(2*(1-\x))});

    \draw[domain=0:1/5, thick, color=blue]
            plot(\x,{0.6*\x/((1-\x))});

    \draw[domain=1/5:2/3, thick, color=blue]
            plot(\x,{0.6*(2*\x+1)/(7*(1-\x))});

    \draw[thick] (0,0)--(1/5,0)--(1/5,1/4*0.6)--(2/3,1/4*0.6)--(2/3,0.6)--(1,0.6);
    \draw[thick] (2/3,0.6)--(1,0.6);
    %\draw[thick] (1/5,1/4*0.6)--(2/3,1/4*0.6);

    \draw[color=red,thick] (0,0)--(2/3,0.6)--(1,0.6);

    %\draw[domain=0:0.5, thick, color=blue]
      %      plot(\x,{0.6*\x/((2-3*\x)}); %C=2

    %\draw[domain=0:0.5, thick, color=green]
         %   plot(\x,{0.6*0.3*\x/((1.3-2.3*\x)}); %C=1.3

    %\draw[thick, color=green] (0,0)--(0.5,0); %C<=1 or cheap talk

    \draw(0,0)node[below left]{0};
    \draw[dashed]  (1,0.6)--(1,0)node[below]{1};
    \draw[dashed]  (2/3,0.6)--(2/3,0)node[below]{$\frac{2}{3}$};
    \draw[dashed]  (2/3,0.6)--(0,0.6)node[left]{1};
    \draw[dashed]  (1/5,1/4*0.6)--(1/5,0)node[below]{$\frac{1}{5}$};

    \end{tikzpicture}
    \caption{Results of \autoref{example:CT=BD=BM}.}
    \label{fig:CT=BD=BM}
\end{figure}

We depict $\mathbb{V}(\mu)$ on \autoref{fig:CT=BD=BM} as the black line, which corresponds to the outcome under a mediator without money burning and cheap talk. Based on \autoref{prop:opt_binary}, we display the result of $\mathcal{V}^*$ on \autoref{fig:CT=BD=BM} as the blue line and $\mathcal{V}^*_{BP}$ as the red line, with the procedure omitted. We observe that, when $\mu_0=\frac{1}{5}$, the case of $\mathcal{V}_{CT}^*(\mu_0)=\mathcal{V}^*_{0}(\mu_0)=\mathcal{V}^*(\mu_0)<\mathcal{V}^*_{BP}(\mu_0)$ arises under \autoref{ass:WeakDominated}.

\section{Omitted Proofs}\label{sec:appendix_proof}

\subsection{Omitted Proofs in \autoref{sec:simplfy}}

\begin{proof}[Proof of \autoref{prop:revelation_principle}]
    For any MDMB $(M,S,\phi)$ and a corresponding PBE assessment $(\sigma,\alpha,\mu)\in\mathcal{E}[\mathcal{G}_{M,S,\phi}(\mu_0)]$, we will directly construct a canonical MDMB $(\pi,x)$ and a corresponding PBE canonical assessment $(\sigma^*,\alpha^*,\mu^*)\in \mathcal{E}[\mathcal{G}_{(\pi,x)}(\mu_0)]$ such that the constructed canonical assessment is payoff-equivalent to the original assessment.

    The canonical MDMB we constructed is as follows: for any $\mu\in\Delta(\Theta),\theta \in \Theta$,
    \begin{equation}\label{eqn:proof_construct_pi}
            \pi(\mu|\theta)=\sum_{s\in S,t\in T,\mu(s,t)=\mu,m\in M}\phi(s,t|m)\sigma(m|\theta),
    \end{equation}
    and for any $\mu\in\Delta(\Theta)$
    \begin{equation}\label{eqn:proof_construct_x}
        x(\mu) =\left\{\begin{matrix}
            \frac{\sum_{s\in S,t\in T, \mu(s,t)=\mu,m\in M,\theta\in\Theta}\phi(s,t|m)\sigma(m|\theta)\mu_0(\theta)t}{\sum_{s\in S,t\in T, \mu(s,t)=\mu,m\in M,\theta\in\Theta}\phi(s,t|m)\sigma(m|\theta)\mu_0(\theta)} & \sum_{s,t, \mu(s,t)=\mu,m,\theta}\phi(s,t|m)\sigma(m|\theta)\mu_0(\theta)\neq 0 \\
            0 & {\rm otherwise}
        \end{matrix}
        \right..
    \end{equation}
    Note that the above canonical MDMB is well-defined since the support of $\phi(m)$ is finite and $x(\mu)\in T$.

    The canonical assessment $(\sigma^*,\alpha^*,\mu^*)$ we constructed is as follows: for all $\theta$, $\sigma^*(\theta|\theta)=1$; for all $\mu\in\Delta(\Theta)$, $\mu^*(\mu) = \mu$; for all $\mu\in\textbf{supp}\{\pi(\theta)\}$ for some $\theta\in\Theta$,
    \begin{equation*}
        \alpha^*(\mu)= \sum_{s\in S,t\in T,\mu(s,t)=\mu}\frac{\sum_{\theta\in\Theta,m\in M}\mu_0(\theta)\sigma(m|\theta)\phi(s,t|m)}{\sum_{s'\in S,t'\geq 0,\mu(s',t')=\mu}\sum_{\theta\in\Theta,m\in M}\mu_0(\theta)\sigma(m|\theta)\phi(s',t'|m)}\alpha(s,t).
    \end{equation*}
    For $\mu\not\in \textbf{supp}\{\pi(\theta)\}$ for any  $\theta\in\Theta$, $\alpha^*(\mu)$ is any best response given posterior belief $\mu$.

    Subsequently, we have to verify two things. The first is that the canonical assessments $(\sigma^*,\alpha^*,\mu^*)\in\mathcal{E}[\mathcal{G}_{\pi,x}(\mu_0)]$. The second is that $(\sigma^*,\alpha^*,\mu^*)$ and $(\sigma,\alpha,\mu)$ are payoff-equivalent.

    Before the verification, we prove the following lemma.
    \begin{lemma}\label{lemma:bayes-equal}
        Suppose $\mu=\mu(s,t)\in\textbf{supp}\{\pi(\hat{\theta})\}$ for some $\hat{\theta}$, then
        $$
        \frac{\sum_{\theta\in\Theta,m\in M}\mu_0(\theta)\sigma(m|\theta)\phi(s,t|m)}{\sum_{\mu(s',t')=\mu}\sum_{\theta\in\Theta,m\in M}\mu_0(\theta)\sigma(m|\theta)\phi(s',t'|m)} = \frac{\sum_{m\in M}\sigma(m|\hat{\theta})\phi(s,t|m)}{\sum_{\mu(s',t')=\mu}\sum_{m\in M}\sigma(m|\hat{\theta})\phi(s',t'|m)}.
        $$
    \end{lemma}
    \begin{proof}[Proof of \autoref{lemma:bayes-equal}]
        By the theorem on equal ratios, it suffices to show that for any $\bar{\theta}\in\textbf{supp}\{\mu\}$, we have that
        \begin{equation}\label{eqn:lemma_bayes-equal}
        \frac{\sum_{m\in M}\sigma(m|\hat{\theta})\phi(s,t|m)}{\sum_{\mu(s',t')=\mu}\sum_{m\in M}\sigma(m|\hat{\theta})\phi(s',t'|m)} = \frac{\sum_{m\in M}\sigma(m|\bar{\theta})\phi(s,t|m)}{\sum_{\mu(s',t')=\mu}\sum_{m\in M}\sigma(m|\bar{\theta})\phi(s',t'|m)}.
        \end{equation}

        According to Bayesian updating, for any $s''\in S,t''\geq 0$ such that $\mu(s,t)=\mu(s'',t'')$, we have that
        $$
        \frac{\mu_0(\hat{\theta})\sum_{m\in M}\sigma(m|\hat{\theta})\phi(s,t|m)}{\mu_0(\bar{\theta})\sum_{m\in M}\sigma(m|\bar{\theta})\phi(s,t|m)}=\frac{\mu(\hat{\theta}|s,t)}{\mu(\bar{\theta}|s,t)}=\frac{\mu(\hat{\theta}|s'',t'')}{\mu(\bar{\theta}|s'',t'')} = \frac{\mu_0(\hat{\theta})\sum_{m\in M}\sigma(m|\hat{\theta})\phi(s'',t''|m)}{\mu_0(\bar{\theta})\sum_{m\in M}\sigma(m|\bar{\theta})\phi(s'',t''|m)}.
        $$
        Thus,
        $$
        \frac{\sum_{m\in M}\sigma(m|\hat{\theta})\phi(s,t|m)}{\sum_{m\in M}\sigma(m|\hat{\theta})\phi(s'',t''|m)} = \frac{\sum_{m\in M}\sigma(m|\bar{\theta})\phi(s,t|m)}{\sum_{m\in M}\sigma(m|\bar{\theta})\phi(s'',t''|m)}.
        $$
        Since $s'',t''$ can be any pair satisfying that $\mu(s'',t'')=\mu$, by the theorem on equal ratios, \autoref{eqn:lemma_bayes-equal} holds.
    \end{proof}

    \textbf{Sender's optimality and payoff-equivalence for the Sender.} To show Sender's optimality and Sender's payoff-equivalence, it is sufficient to show that the expected payoffs of type $\theta$ Sender under both assessments are the same. The expected payoff of type $\theta$ Sender under the assessment $(\sigma,\alpha,\mu)$ is
    $$
    \sum_{m\in M,s\in S, t\in T, a\in A}\sigma(m|\theta)\phi(s,t|m)\alpha(a|s,t)(v(a)-t).
    $$
    The expected payoff of type $\theta$ Sender under the assessment $(\sigma^*,\alpha^*,\mu^*)$ is
    $$
    \sum_{\mu\in\textbf{supp}\{\pi(\theta)\}, a\in A}\pi(\mu|\theta)\alpha^*(a|\mu)(v(a)-x(\mu)).
    $$
    These two expected payoffs are the same, since by \autoref{lemma:bayes-equal}, for any $a\in A,s\in S,t\in T$ the coefficient of $\alpha(a|s,t)$ where $\mu(s,t)=\mu\in\textbf{supp}\{\pi(\theta)\}$ in the expression $$\sum_{\mu\in\textbf{supp}\{\pi(\theta)\}, a\in A}\pi(\mu|\theta)\alpha^*(a|\mu)v(a)$$ is
    \begin{equation*}
        \begin{split}
            & \sum_{s',t',m,\mu(s',t')=\mu} \sigma(m|\theta)\phi(s',t'|m)\frac{\sum_{\theta'\in\Theta,m\in M}\mu_0(\theta')\sigma(m|\theta')\phi(s,t|m)}{\sum_{s'\in S,t'\geq 0,\mu(s',t')=\mu}\sum_{\theta'\in\Theta,m\in M}\mu_0(\theta')\sigma(m|\theta')\phi(s',t'|m)}v(a) \\
            & =\sum_{s',t',m,\mu(s',t')=\mu} \sigma(m|\theta)\phi(s',t'|m)\frac{\sum_{m}\sigma(m|\theta)\phi(s,t|m)}{\sum_{s',t',m,\mu(s',t')=\mu} \sigma(m|\theta)\phi(s',t'|m)}v(a) \\
            & =\sum_{m}\sigma(m|\theta)\phi(s,t|m)v(a)
        \end{split}
    \end{equation*}
    where the second equation holds by \autoref{lemma:bayes-equal}, and expected money burning of type-$\theta$ Sender of $(\sigma^*,\alpha^*,\mu^*)$ is
    \begin{equation*}
    \begin{split}
            \sum_{\mu\in\textbf{supp}\{\pi(\theta)\}, a\in A}\pi(\mu|\theta)\alpha^*(a|\mu)x(\mu) & =  \sum_{\mu\in\textbf{supp}\{\pi(\theta)\}}\pi(\mu|\theta)x(\mu)  \\
            & = \sum_{\mu\in\textbf{supp}\{\pi(\theta)\}}\pi(\mu|\theta)\frac{\sum_{s,t,\mu(s,t)=\mu,m\in M}\sigma(m|\theta)\phi(s,t|m)t}{\sum_{\mu(s,t)=\mu}\sum_{m\in M}\sigma(m|\theta)\phi(s,t|m)}\\
            & = \sum_{m\in M,s\in S, t\in T}\sigma(m|\theta)\phi(s,t|m)t,
    \end{split}
    \end{equation*}
    where the second equation holds by \autoref{lemma:bayes-equal}.

    \textbf{Receiver's optimality.} 
    Since $\alpha(s,t)$ is the best response under the belief $\mu(s,t)$ and $\alpha^*(\mu)$ is a convex combination of some $\alpha(s',t')$ where $\mu(s',t')=\mu$, by the convexity of the best response set, $\alpha^*(\mu)$ must satisfy the Receiver's optimality condition.

    \textbf{Bayesian updating.} Given $\mu\in\Delta(\Theta)$, for any $s,t$ such that $\mu(s,t)=\mu=\mu^*(\mu)$, by Bayesian updating, we have that for any $\theta\in\Theta$
    $$
    \mu(\theta|s,t)\sum_{\theta'\in\Theta,m\in M}\mu_0(\theta')\sigma(m|\theta')\phi(s,t|m)=\mu_0(\theta)\sum_{m\in M}\sigma(m|\theta)\phi(s,t|m).
    $$
    Hence,
    $$
    \mu(\theta)\sum_{s,t,\mu(s,t)=\mu}\sum_{\theta'\in\Theta,m\in M}\mu_0(\theta')\sigma(m|\theta')\phi(s,t|m)=\sum_{s,t,\mu(s,t)=\mu}\mu_0(\theta)\sum_{m\in M}\sigma(m|\theta)\phi(s,t|m).
    $$
    That is
    $$
\mu^*(\theta|\mu)\sum_{\theta'}\mu_0(\theta')\pi(\mu|\theta')=\mu_0(\theta)\pi(\mu|\theta).
    $$
    \textbf{Receiver's payoff-equivalence.} Receiver's payoff under assessment $(\sigma,\alpha,\mu)$ is 
    \[
    \sum_{\theta\in\Theta}\mu_0(\theta)\sum_{m,s,t,a}\sigma(m|\theta)\phi(s,t|m)\alpha(a|s,t)u(a,\theta).
    \]
    Receiver's payoff under assessment $(\sigma^*,\alpha^*,\mu^*)$ is 
    \[
    \sum_{\theta\in\Theta}\mu_0(\theta)\sum_{\mu\in\textbf{supp}\{\pi(\theta)\}}\sum_{a\in A}\pi(\mu|\theta)\alpha^*(a|\mu)u(a,\theta).
    \]
    These two payoffs are the same since, by the definition of $\alpha^*$ and \autoref{lemma:bayes-equal},
    \begin{equation*}
        \begin{split}
            &\sum_{\theta\in\Theta}\mu_0(\theta)\sum_{\mu\in\textbf{supp}\{\pi(\theta)\}}\sum_{a\in A}\pi(\mu|\theta)\alpha^*(a|\mu)u(a,\theta)\\
            &=\sum_{\theta\in\Theta}\mu_0(\theta)\sum_{\mu\in\textbf{supp}\{\pi(\theta)\}}
            \sum_{\substack{s\in S,t\in T:\\ \mu(s,t)=\mu}}\sum_{a\in A}
            \pi(\mu|\theta)
            \frac{\sum_{\theta'\in\Theta,m\in M}\mu_0(\theta')\sigma(m|\theta')\phi(s,t|m)}
            {\sum_{\substack{s'\in S,t'\in T:\\ \mu(s',t')=\mu}}\sum_{\theta'\in\Theta,m\in M}\mu_0(\theta')\sigma(m|\theta')\phi(s',t'|m)}
            \alpha(a|s,t)u(a,\theta) \\
            &=\sum_{\theta\in\Theta}\mu_0(\theta)\sum_{\mu\in\textbf{supp}\{\pi(\theta)\}}
            \sum_{\substack{s\in S,t\in T:\\ \mu(s,t)=\mu}}\sum_{a\in A}\sum_{m\in M}
            \sigma(m|\theta)\phi(s,t|m)\alpha(a|s,t)u(a,\theta) \\
            &=\sum_{\theta\in\Theta}\mu_0(\theta)\sum_{m,s,t,a}\sigma(m|\theta)\phi(s,t|m)\alpha(a|s,t)u(a,\theta).
        \end{split}
    \end{equation*}
\end{proof}

\subsection{Omitted Proofs in \autoref{sec:upper_bound}}

\begin{proof}[Proof of \autoref{thm:SenderOPT_bounded}]
According to \autoref{coro:simply_value_opt}, we need to solve the following optimization problem.
\begin{alignat}{2}
    \max \quad & k & {} & \label{prg:SOP_bound}\\
    \mbox{s.t} \quad & k=\int_{\mu}\frac{\mu(\theta)}{\mu_0(\theta)}(V(\mu)-x(\mu))dp(\mu)& \quad & \forall \theta\in \Theta \tag{IC} \\
    & p\in BP(\mu_0) &\quad &  \tag{BP} \\
    & V(\mu) \in \mathbb{V}(\mu) & \quad & \forall \mu\in \Delta(\Theta)\tag{O} \\
    & 0\leq x(\mu)\leq C &\quad & \forall \mu\in \Delta(\Theta) \tag{Bgt}
\end{alignat}

We adopt a two-step optimization approach. First, fixing the signaling scheme $p\in BP(\mu_0)$, we try to find the optimal burning scheme $x(\mu)$ where $0\leq x(\mu)\leq C$ and the Receiver's best response $V(\mu)\in\mathbb{V}(\mu)$. Now, it is a linear programming problem. By the fundamental theorem of linear programming, we can also obtain $\mathcal{V}^*_{C}(\mu_0)$ from the following max-min problem.
\[
    \max_{p\in BP(\mu_0)}\min_{\lambda\in \text{aff}(\Theta)} \int_{\mu}\max\{\sum_{\theta\in\Theta}\lambda(\theta)\frac{\mu(\theta)}{\mu_{0}(\theta)}\max \mathbb{V}(\mu),\sum_{\theta\in\Theta}\lambda(\theta)\frac{\mu(\theta)}{\mu_{0}(\theta)} (\min \mathbb{V}(\mu)-C)\}dp(\mu).
\]
This implies that we choose $V(\mu)=\max \mathbb{V}(\mu),x(\mu)=0$ if $\sum_{\theta\in\Theta}\lambda(\theta)\frac{\mu(\theta)}{\mu_{0}(\theta)}> 0$ and $V(\mu)=\min \mathbb{V}(\mu),x(\mu)=C$ if $\sum_{\theta\in\Theta}\lambda(\theta)\frac{\mu(\theta)}{\mu_{0}(\theta)}< 0$, which determines the choice of the receiver's best responses.

The rest of the proof relies on Sion's minimax theorem as well. It is easy to verify that $BP(\mu_0)$ is a compact and convex set, and $\{\lambda|\lambda\in \text{aff}(\Theta)\}$ is a convex set. Moreover, $\int_{\mu}\max\{\sum_{\theta\in\Theta}\lambda(\theta)\frac{\mu(\theta)}{\mu_{0}(\theta)}\max \mathbb{V}(\mu),\sum_{\theta\in\Theta}\lambda(\theta)\frac{\mu(\theta)}{\mu_{0}(\theta)} (\min \mathbb{V}(\mu)-C)\}dp(\mu)$ is linear in $p$ and continuous convex in $\lambda$ since it is the maximum of two linear functions. Since $\max\mathbb V(\mu)$ and $\min\mathbb V(\mu)-C$ are upper and lower semi-continuous, respectively, we have that $\sum_{\theta\in\Theta}\lambda(\theta)\frac{\mu(\theta)}{\mu_{0}(\theta)}\max\mathbb V(\mu)$ is upper semi-continuous when $\sum_{\theta\in\Theta}\lambda(\theta)\frac{\mu(\theta)}{\mu_{0}(\theta)}>0$ and $\sum_{\theta\in\Theta}\lambda(\theta)\frac{\mu(\theta)}{\mu_{0}(\theta)}(\min\mathbb V(\mu)-C)$ is upper semi-continuous when $\sum_{\theta\in\Theta}\lambda(\theta)\frac{\mu(\theta)}{\mu_{0}(\theta)}<0$. Therefore, $$\hat{V}_{\lambda,C}(\mu)=\max\{\sum_{\theta\in\Theta}\lambda(\theta)\frac{\mu(\theta)}{\mu_{0}(\theta)}\max\mathbb V(\mu),\sum_{\theta\in\Theta}\lambda(\theta)\frac{\mu(\theta)}{\mu_{0}(\theta)} (\min\mathbb V(\mu)-C)\}$$ is upper semi-continuous and so is $\int_{\mu}\max\{\sum_{\theta\in\Theta}\lambda(\theta)\frac{\mu(\theta)}{\mu_{0}(\theta)}\max \mathbb{V}(\mu),\sum_{\theta\in\Theta}\lambda(\theta)\frac{\mu(\theta)}{\mu_{0}(\theta)} (\min \mathbb{V}(\mu)-C)\}dp(\mu)$ in $p$. Hence, we can apply Sion's minimax theorem directly and complete the proof.
\end{proof}

\subsection{Omitted Proofs in \autoref{sec:improvement}}
\begin{proof}[Proof of \autoref{thm:credibility_gaining_group}]
    Without loss of generality, we suppose $v(a)> 0$ for all $a\in A$ and $C_2>C_1>\max_a v(a)$. In the following proof, we prove that either $\mathcal{V}_{C_2}^*(\mu_0)>\mathcal{V}_{C_1}^*(\mu_0)$ or $\mathcal{V}_{C_2}^*(\mu_0)=\mathcal{V}_{C_1}^*(\mu_0)=\mathcal{V}^*_{0}(\mu_0)=\mathcal{V}^*_{CT}(\mu_0)$. 

     We adopt a proof by contradiction. Let $\mu_0$ be a belief with the \textbf{smallest} support such that $\mathcal{V}_{CT}^*(\mu_0)<\mathcal{V}^*_{C_1}(\mu_0)=\mathcal{V}_{C_2}^*(\mu_0)$ where $C_2>C_1>\max_a v(a)$. Let $(p_{C_1},x_{C_1},V_{C_1})$ denote the optimal solution to the program \eqref{eqn:SenderOPT_Canonical} given the budget $C_1$ and let $\lambda_{C_1}$ denote the corresponding Lagrange multiplier. Since $\mathcal{V}^*_{C_2}(\mu_0)=\mathcal{V}_{C_1}^*(\mu_0)$, $(p_{C_1},x_{C_1},V_{C_1})$ is also optimal to the program \eqref{eqn:SenderOPT_Canonical} given the budget $C_2$ and let $\lambda_{C_2}$ denote the corresponding Lagrange multiplier. Because $x_{C_1}(\mu)\leq C_1<C_2$, by \autoref{prop:opt_MDMB} (ii), we know that for any $\mu\in\textbf{supp}\{p_{C_1}\}$, $\sum_{\theta\in\Theta}\lambda_{C_2}(\theta)\frac{\mu(\theta)}{\mu_0(\theta)}\geq 0$. Now, let $U_1=\{\mu\in\textbf{supp}\{p_{C_1}\}\big| \sum_{\theta\in\Theta}\lambda_{C_2}(\theta)\frac{\mu(\theta)}{\mu_0(\theta)}>0\}$ and $U_2=\{\mu\in\textbf{supp}\{p_{C_1}\}\big| \sum_{\theta\in\Theta}\lambda_{C_2}(\theta)\frac{\mu(\theta)}{\mu_0(\theta)}=0\}$. We know that $U_1$ and $U_2$ are a partition of $\textbf{supp}\{p_{C_1}\}$. Next, we prove two claims. 
    \begin{claim}\label{cl:persuasion_to_CT}
        For any $\mu,\mu'\in U_1$, $\max \mathbb{V}(\mu)=\max \mathbb{V}(\mu')$.
    \end{claim}
    \begin{proof}[Proof of \autoref{cl:persuasion_to_CT}]
            By the optimality of $(p_{C_1},x_{C_1},V_{C_1})$ and $\lambda_{C_2}$, we have that
$$
\int_{\mu}\frac{\mu(\theta)}{\mu_0(\theta)} (V_{C_1}(\mu)-x_{C_1}(\mu))d p_{C_1}=\int_{\mu}\sum_{\theta}\frac{\lambda_{C_2}(\theta)\mu(\theta)}{\mu_0(\theta)} (V_{C_1}(\mu)-x_{C_1}(\mu)) d p_{C_1}(\mu),
$$
and $p_{C_1}$ is the concavification of $\hat{V}_{\lambda_{C_2},C_2}$ at $\mu_0$. Then, by Proposition 9 of the working paper version of \cite{kamenica2011bayesian}, we have that there exist parameters $A_{\theta}$ for $\theta\in\Theta$ and for any $\mu\in\textbf{supp}\{p_{C_1}\}$,
$$
\sum_{\theta}\frac{\lambda_{C_2}(\theta)\mu(\theta)}{\mu_0(\theta)} (V_{C_1}(\mu)-x_{C_1}(\mu))=\sum_{\theta}A_{\theta}\mu(\theta).
$$
Hence, we obtain that for any $\theta\in\Theta$,
$$
A_{\theta}\mu_0(\theta)\int_{\mu}\frac{\mu(\theta)}{\mu_0(\theta)} (V_{C_1}(\mu)-x_{C_1}(\mu))d p_{C_1}(\mu) = A_{\theta}\mu_0(\theta)\sum_{\theta'}A_{\theta'}\mu_0(\theta').
$$
Summing them up and since $0=\sum_{\theta\in\Theta}\lambda_{C_2}(\theta)\frac{\mu(\theta)}{\mu_0(\theta)}=\sum_{\theta\in\Theta}A_\theta\mu(\theta)$ for any $\mu\in U_2$, we have that
$$
\int_{\mu\in U_1}\sum_{\theta} A_{\theta}\mu(\theta) \frac{\sum_{\theta} A_{\theta}\mu(\theta)}{\sum_{\theta}\frac{\lambda_{C_2}(\theta)\mu(\theta)}{\mu_0(\theta)}} d p_{C_1}(\mu) = (\sum_{\theta}A_{\theta}\mu_0(\theta))^2.
$$
Since $\int_{\mu\in U_1}\sum_{\theta}\frac{\lambda_{C_2}(\theta)\mu(\theta)}{\mu_0(\theta)}d p_{C_1}(\mu)=\int_{\mu}\sum_{\theta}\frac{\lambda_{C_2}(\theta)\mu(\theta)}{\mu_0(\theta)}d p_{C_1}(\mu)=1$, by the equality condition of Cauchy's inequality and the above equation, we have that for any $\mu,\mu'\in U_1$,
$$
\frac{\sum_{\theta} A_{\theta}\mu(\theta)}{\sum_{\theta}\frac{\lambda_{C_2}(\theta)\mu(\theta)}{\mu_0(\theta)}}=\frac{\sum_{\theta} A_{\theta}\mu'(\theta)}{\sum_{\theta}\frac{\lambda_{C_2}(\theta)\mu'(\theta)}{\mu_0(\theta)}}.
$$
By \autoref{prop:opt_MDMB}, this implies that $ \max \mathbb V(\mu)= \max \mathbb V (\mu')$ for $\mu,\mu'\in U_1$. To simplify the notation, we set this value to be $R$, which coincides with $\mathcal{V}_{C_1}^*(\mu_0)$ and $\mathcal{V}_{C_2}^*(\mu_0)$.
\end{proof}

    \begin{claim}\label{cl:balance_boundary}
        For any $\theta \in \textbf{supp}\{\mu\},\mu\in U_2$, $\lambda_{C_2}(\theta) = 0$. 
    \end{claim}
    \begin{proof}[Proof of \autoref{cl:balance_boundary}]
        Let $q_1=\int_{\mu\in U_1} dp_{C_1}(\mu), q_2=\int_{\mu\in U_2} dp_{C_1}(\mu)$ and $\mu_1=\int_{\mu\in U_1} \mu dp_{C_1}(\mu)/q_1, \mu_2=\int_{\mu\in U_2} \mu dp_{C_1}(\mu)/q_2$. We only have to show that for any $\theta\in\textbf{supp}\{\mu_2\}$, $\lambda_{C_2}(\theta)=0$.

        We prove by contradiction. Suppose there is $\theta\in\textbf{supp}\{\mu_2\}$, $\lambda_{C_2}(\theta)\neq 0$, thus there is $\hat{\theta}\in \textbf{supp}\{\mu_2\}$ such that $\lambda_{C_2}(\hat{\theta})<0$. Therefore,
        \[
        \sum_{\theta\in \textbf{supp}\{\mu_2\}} \hat{V}_{\lambda_{C_2},C_2}(\mu_\theta) \mu_2(\theta) \geq \mu_2(\hat{\theta})\lambda_{C_2}(\hat{\theta}) (\min \mathbb{V}(\mu_{\hat{\theta}})-C_2) > 0 =\int_{\mu\in U_2}\hat{V}_{\lambda_{C_2},C_2}(\mu) dp_{C_1}(\mu) 
        \]
        where the last inequality is because $C_2>\max_{a}v(a)$. This contradicts to the fact that $p_{C_1}$ is the concavification of $\hat{V}_{\lambda_{C_2},C_2}$.
    \end{proof}

Now, back to our proof. For any $\theta\in\textbf{supp}\{\mu_2\}$, we have that
$$
R=\int_{\mu}\frac{\mu(\theta)}{\mu_0(\theta)}(V_{C_1}(\mu)-x_{C_1}(\mu))dp_{C_1}(\mu)=q_1 \frac{\mu_1(\theta)}{\mu_0(\theta)}R + q_2\frac{\mu_2(\theta)}{\mu_{0}(\theta)}\int_{\mu\in U_2}\frac{\mu(\theta)}{\mu_2(\theta)}(V_{C_1}(\mu)-x_{C_1}(\mu))d\frac{p_{C_1}(\mu)}{q_2}.
$$
Hence, for any $\theta\in \textbf{supp}\{\mu_2\}$,
$$
\int_{\mu\in U_2}\frac{\mu(\theta)}{\mu_2(\theta)}(V_{C_1}(\mu)-x_{C_1}(\mu))d\frac{p_{C_1}(\mu)}{q_2} = R.
$$
This implies that $\mathcal{V}_{C_1}^*(\mu_2)\geq R$. Next, we divide the rest of the proof into two cases. 

\textbf{Case 1}: Suppose that $\mathcal{V}^*_{C_2}(\mu_2)=R$. Since we have already shown that $\mathcal{V}^*_{C_1}(\mu_2)\geq R$, and since $\mathcal{V}^*_{C_2}(\mu_2)\geq \mathcal{V}^*_{C_1}(\mu_2)$ by monotonicity in the budget, it follows that
\[
\mathcal{V}^*_{C_2}(\mu_2)=\mathcal{V}^*_{C_1}(\mu_2)=R.
\]
Moreover, $|\textbf{supp}\{\mu_2\}|<|\textbf{supp}\{\mu_0\}|$. Otherwise, by \autoref{cl:balance_boundary}, $\lambda_{C_2}(\theta)=0$ for all $\theta\in\textbf{supp}\{\mu_0\}$, which would imply $\mathcal{V}^*_{C_2}(\mu_0)=0$, a contradiction. Hence, by the minimal choice of $\mu_0$ as a smallest-support counterexample, we must have $\mathcal{V}_{CT}^*(\mu_2)=R$.

Let $p_2\in BP(\mu_2)$ be a cheap-talk distribution that attains $\mathcal{V}_{CT}^*(\mu_2)=R$. Combining $p_2$ with the conditional distribution $p_{C_1}(\cdot\mid U_1)$ using weights $q_2$ and $q_1$ yields a Bayes-plausible distribution for $\mu_0$, because $q_1\mu_1+q_2\mu_2=\mu_0$. For posteriors in $U_1$, we have $\max \mathbb V(\mu)=R$ by \autoref{cl:persuasion_to_CT}; for posteriors induced by $p_2$, the Sender can obtain payoff at least $R$ by the cheap-talk construction for $\mu_2$. Therefore, by \cite{lipnowski2020cheap}, there exists a cheap-talk equilibrium at prior $\mu_0$ in which the Sender obtains payoff at least $R$. Thus $\mathcal{V}_{CT}^*(\mu_0)\geq R$, contradicting the assumption that $\mathcal{V}_{CT}^*(\mu_0)<R$.

\textbf{Case 2}: Suppose that $\mathcal{V}^*_{C_2}(\mu_2)>R$. Then, there must exist $\mu_3$ where $\textbf{supp}\{\mu_3\}\subseteq \textbf{supp}\{\mu_2\}$ such that $\max \mathbb{V}(\mu_3) > R$. By \autoref{ass:WeakDominated}, we choose $\mu_3$ such that $RO(\mu_3)$ is a singleton. So, we can select small enough $\varepsilon>0$ such that $\max \mathbb{V}(\hat{\mu}_3) = \max \mathbb{V}(\mu_3)$ where $\hat{\mu}_3 = (1-\varepsilon)\mu_3+\varepsilon\mu_1$. On the one hand, we have 
\begin{equation*}
    \begin{split}
        \hat{V}_{\lambda_{C_2},C_2}(\hat{\mu}_3)& = \sum_{\theta\in\Theta}\lambda_{C_2}(\theta)\frac{(1-\varepsilon)\mu_3(\theta)+\varepsilon\mu_1(\theta)}{\mu_0(\theta)} \max \mathbb V(\mu_3) \\
        & =\varepsilon\sum_{\theta\in\Theta}\lambda_{C_2}(\theta)\frac{\mu_1(\theta)}{\mu_0(\theta)} \max \mathbb V(\mu_3) \\
        & > \varepsilon \sum_{\theta}\lambda_{C_2}(\theta)\frac{\mu_1(\theta)}{\mu_0(\theta)} R.
    \end{split}
\end{equation*}
On the other hand, because $p_{C_1}$ is the concavification of $\hat{V}_{\lambda_{C_2},C_2}$, we have 
\begin{equation*}
    \begin{split}
        \hat{V}_{\lambda_{C_2},C_2}(\hat{\mu}_3) \leq \sum_{\theta} A_{\theta}\hat{\mu}_3(\theta)  =\varepsilon\sum_{\theta}A_{\theta}\mu_1(\theta)& =\varepsilon\int_{\mu\in U_1}\sum_{\theta\in\Theta}\lambda_{C_2}(\theta)\frac{\mu(\theta)}{\mu_0(\theta)}\max \mathbb V(\mu)d\frac{p_{C_1}(\mu)}{q_1} \\
        & = \varepsilon \sum_{\theta}\lambda_{C_2}(\theta)\frac{\mu_1(\theta)}{\mu_0(\theta)} R
    \end{split}
\end{equation*}
where the last equation is by \autoref{cl:persuasion_to_CT}. So far, we have obtained a contradiction, and we complete the proof. 
\end{proof}

\begin{proof}[Proof of \autoref{prop:improve_CT_C}]
Let $\bar{\mu}$ be a belief that attains $\max_{\mu\in\Delta(\Theta)}\max\mathbb V(\mu)$, and let $a\in RO(\bar{\mu})$ be an action that delivers this value to the Sender. By \autoref{ass:WeakDominated}, there exists $\tilde{\mu}$ with $\textbf{supp}\{\tilde{\mu}\}=\textbf{supp}\{\bar{\mu}\}$ such that $RO(\tilde{\mu})=\{a\}$. Since the model is discrete, both $\Theta$ and $A$ are finite, and the Receiver's payoff from each action is continuous and linear in the belief. Hence the strict optimality of $a$ at $\tilde{\mu}$ is preserved under all sufficiently small perturbations of $\tilde{\mu}$. Therefore, we can choose a full-support belief $\hat{\mu}$ sufficiently close to $\tilde{\mu}$ such that $RO(\hat{\mu})=\{a\}$, which implies $\max\mathbb V(\hat{\mu})=\max_{\mu\in\Delta(\Theta)}\max\mathbb V(\mu)$.

Using the terminology from \cite{corrao2023communication}, since there is a sufficiently small $\varepsilon>0$ such that $qcav(\mathbb V)(\mu_0+\varepsilon(\mu-\mu_0))=qcav(\mathbb V)(\mu_0)$ for all $\mu\in \Delta(\Theta)$, then $\mu_0$ satisfies the full-dimensionality condition; and since $C>\max_{\mu,V(\mu)\in\mathbb{V}(\mu)} V(\mu)-\min_{\mu,V(\mu)\in\mathbb{V}(\mu)} V(\mu)$, for $\hat{\mu}'=(1-t)\mu_0 + t\hat{\mu}$ where $t>1$, the Sender can induce a payoff $\min \mathbb V(\hat{\mu}')-C<qcav(\mathbb V)(\mu_0)$ at $\hat{\mu}'$. Therefore, the prior $\mu_0$ is also locally improvable under the budget $C$. Thus, according to Theorem 3 in \cite{corrao2023communication}, we have that $\mathcal{V}^*_{C}(\mu_0)>\mathcal{V}^*_{CT}(\mu_0)$. Hence, $\mathcal{V}^*_{C}(\mu_0)>\mathcal{V}^*_{0}(\mu_0)$ by \autoref{thm:credibility_gaining_group}.
\end{proof}

\subsection{Omitted Proofs in \autoref{sec:value_MDMB}}

\begin{proof}[Proof of \autoref{cl:construct_noise}]
Given that $V_{\bar{\pi}}(\theta)-\min_{\theta'\in\Theta}V_{\bar{\pi}}(\theta')\geq 0$, it follows that $\bar{x}(\mu)\geq 0$ for all $\mu\in \Delta(\Theta)$. Therefore, for any $\theta\in\Theta$, the Sender's payoff of type $\theta$ under the mechanism $(\bar{\pi},\bar{x})$ is equal to
$$
V_{\bar{\pi}}(\theta)-\sum_{\mu\in\textbf{supp}\{\bar{\pi}(\theta)\}}\bar{\pi}(\mu|\theta) \bar{x}(\mu)=\min_{\theta'\in\Theta}V_{\bar{\pi}}(\theta').
$$
This implies that the mechanism is incentive-compatible. Moreover, the Sender's expected payoff is $\min_{\theta\in\Theta}V_{\bar{\pi}}(\theta)$.

Next, we show that the corresponding canonical assessments also satisfy the Bayesian updating condition under this MDMB mechanism. Let $\mu_{\theta}\in \Delta(\Theta)$ be such that $\mu_{\theta}(\theta)=1$ and $\mu_{\theta}(\theta')=0$ for any $\theta'\neq \theta$. For any non-degenerate posterior $\mu$, the transition probabilities are scaled by the same factor $1-\delta$ across types, so the induced posterior remains $\mu$. For the degenerate posterior $\mu_{\theta}$, Bayes rule assigns probability one to type $\theta$, because $\bar{\pi}(\mu_{\theta}|\theta)>0$ and $\bar{\pi}(\mu_{\theta}|\theta')=0$ for all $\theta'\neq\theta$, where the latter follows from Bayes plausibility of $\pi$ and the full-support prior. Thus the construction is canonical. Hence, $V_{\bar{\pi}}(\theta)=(1-\delta)V_{\pi}(\theta)+\delta\max\mathbb V(\mu_{\theta})$. It follows that
$$
\lim_{\delta\rightarrow 0^+}\min_{\theta\in\Theta}\{V_{\bar{\pi}}(\theta)\} = \lim_{\delta\rightarrow 0^+}\min_{\theta\in\Theta}\{(1-\delta)V_{\pi}(\theta)+\delta\max\mathbb V(\mu_{\theta})\} = \min_{\theta \in \Theta}V_{\pi}(\theta).
$$
\end{proof}

\section{Binary Types Cases}
In this section, we assume that the Sender has only two possible types, $\Theta=\{\theta_1,\theta_2\}$. In this case, we can give a clearer characterization of the persuasion group and the credibility-gaining group.

First, the following proposition states that the worst subjective prior must be one of the extreme points of $\Delta(\Theta)$. Thus, when we search for the persuasion group if the budget $C$ is sufficiently large, we only need to consider the concavification over $\hat{V}_{\theta_1}$ or $\hat{V}_{\theta_2}$. 
\begin{proposition}\label{prop:opt_binary}
Suppose the Sender has a binary type set $\Theta=\{\theta_1,\theta_2\}$. Then, for any prior distribution $\mu_0\in\Delta(\Theta)$, we have $\mathcal{V}^*(\mu_0)=\min\{cav(\hat{V}_{\theta_1})(\mu_0),cav(\hat{V}_{\theta_2})(\mu_0)\}$.
\end{proposition}
\begin{proof}[Proof of \autoref{prop:opt_binary}]
By \autoref{prop:opt_lambda}, it suffices to show that there exists $\theta_i\in\{\theta_1,\theta_2\}$ such that for any $p\in BP(\mu_0)$ with $\mathcal{L}(\theta_i,p)=cav(\hat{V}_{\theta_i})(\mu_0)$, we have $\mathcal{L}(\theta_i,p)\leq \mathcal{L}(\theta_{3-i},p)$.

Let $U=\{\mu|V(\mu)=\max_{x\in[0,1]} V(x)\}$ denote the range of posteriors that yield the maximum value for Sender, for $\mu\in[0,1]$. Since $U$ is closed and $V(\cdot)$ is upper semi-continuous, $U$ can be expressed as the union of closed intervals. We assume that $l=\min U$ and $r=\max U$. If $l\leq \mu_0\leq r$, then it is clear that $V^*(\mu_0)=\max_{x\in[0,1]} V(x)$ and for any $\theta_i$, $cav(\hat{V}_{\theta_i})(\mu_0)=V^*(\mu_0)$, which is a constant. Hence, our statement holds trivially. In the following proof, we consider $\mu_0>r$ or $\mu_0<l$.

Without loss of generality, we assume that $\mu_0>r\geq 0$ by symmetry. We focus on $\theta_1$. We prove by contradiction that if there exists $p\in BP(\mu_0)$ with $\mathcal{L}(\theta_1,p)=cav(\hat{V}_{\theta_1})(\mu_0)$ and $\mathcal{L}(\theta_1,p)> \mathcal{L}(\theta_{2},p)$, then we reach a contradiction. Since $p\in BP(\mu_0)$ performs the concavification of the function $\hat{V}_{\theta_1}$ at point $\mu_0$, by Proposition 9 of the working paper version of \cite{kamenica2011bayesian}, we have that the points $(\mu,\hat{V}_{\theta_1}(\mu))$ for $\mu\in\textbf{supp}\{p\}$ are collinear. This means that there exist parameters $k,b$ such that for any  $\mu\in\textbf{supp}\{p\}$,
 $$
 \frac{\mu}{\mu_0} V(\mu) = k\mu + b.
 $$
 Since $\mathcal{L}(\theta_1,p)> \mathcal{L}(\theta_{2},p)$, we have
 $$
 \int_{\mu}(\mu-\mu_0) V(\mu)dp(\mu)>0.
 $$
 Substituting $V(\mu)$ with $\mu_0(k+\frac{b}{\mu})$ and using $\int_{\mu}(\mu-\mu_0)dp(\mu)=0$, we obtain
 $$
 b(1-\int_{\mu}\frac{\mu_0}{\mu}dp(\mu))=\int_{\mu}(\mu-\mu_0)\frac{b}{\mu}dp(\mu)>0.
 $$
 By Cauchy's inequality,
$$
\int_{\mu}\frac{\mu_0}{\mu}dp(\mu)=\int_{\mu}\frac{\mu}{\mu_0}dp(\mu)\int_{\mu}\frac{\mu_0}{\mu}dp(\mu)\geq (\int_{\mu}dp(\mu))^2=1.
$$
Therefore, we must have $b<0$. However, $k\mu+b$ is the concavification line of $\hat{V}_{\theta_1}(\cdot)$ at $\mu_0$. Thus, it must satisfy that for any $\mu\in[0,1]$, $\hat{V}_{\theta_1}(\mu)\leq k\mu+b$. Choosing $\mu=0$, we get $b\geq 0$. This is a contradiction.
\end{proof}

In \autoref{example:regulationPD}, however, we show that the worst subjective prior is not necessarily an extreme point of $\Delta(\Theta)$ in general. 

\begin{example}\label{example:regulationPD}
We consider an example with three parties: a seller, a buyer, and an influencer. The seller wants to sell a zero-cost product to the buyer. The buyer's valuation of the product is $v$. The seller only knows that $v$ is distributed uniformly in $\{1,2,3\}$. The buyer is a fan of the influencer, who wants to help the buyer reduce the price of the product by disclosing information about the buyer's type and subsidizing the seller. The influencer acts as the Sender who uses our MDMB to influence the seller's action as the Receiver. To fit our model, we let the type set be $\Theta=\{v_1=1,v_2=2,v_3=3\}$, the prior distribution be $\mu_{0}=(\frac{1}{3},\frac{1}{3},\frac{1}{3})$, and the seller's action set be $A=\{p_1=1,p_2=2,p_3=3\}$.

We assume that the influencer's objective is to minimize the price of the product. If the seller charges a price $p_i$ to the buyer, the influencer's valuation function is $v(p_i)=4-p_i$.
\end{example}
We use \autoref{thm:min_max} to examine the extreme point subjective priors of $\Delta(\Theta)$ at first. We then use \autoref{prop:opt_lambda} to find the worst Sender's subjective prior and the corresponding maximum payoff of the influencer achieved by the MDMB. 

We consider three extreme point subjective priors $\lambda_i\in\Delta(\Theta),i=1,2,3$, where $\lambda_1=(1,0,0),\lambda_2=(0,1,0),\lambda_3=(0,0,1)$. For any $\lambda\in\Delta(\Theta)$, to find the concavification value of $\hat{V}_{\lambda}(\mu)$ at $\mu_0$, we can assume without loss of generality that we only need to find the distribution of posterior $\tau\in BP(\mu_0)$ that induces different actions of the Receiver.\footnote{This is true because if two posteriors $\mu_1,\mu_2$ in the support of $\tau$ lead to the same action of the Receiver, we can merge them as posterior $\frac{\tau(\mu_1)}{\tau(\mu_1)+\tau(\mu_2)}\mu_1+\frac{\tau(\mu_2)}{\tau(\mu_1)+\tau(\mu_2)}\mu_2$ with probability $\tau(\mu_1)+\tau(\mu_2)$.} Then finding $cav(\hat{V}_{\lambda})(\mu_0)$ becomes a linear programming problem. We obtain that $cav(\hat{V}_{\lambda_1})(\mu_0)=3$, where $\{(1,0,0),(0,1,0),(0,0,1)\}$ form the support of Receiver's posterior distribution and they are realized with equal probability; $cav(\hat{V}_{\lambda_2})(\mu_0)=3$, where $(\frac{1}{2},\frac{1}{2},0),(0,0,1)$ form the support of Receiver's posterior distribution and they are realized with probability $\frac{2}{3},\frac{1}{3}$, respectively; and $cav(\hat{V}_{\lambda_3})(\mu_0)=\frac{8}{3}$. So we can conclude that in this example $\mathcal{V}^*(\mu_0)\leq \frac{8}{3}$.

However, $\lambda_3$ is not the worst subjective prior in this case, even though it minimizes $cav(\hat{V}_{\lambda})(\mu_0)$ among the extreme points of $\Delta(\Theta)$. Next, we show that $\lambda^*=(0,\frac{1}{2},\frac{1}{2})$ is the worst subjective prior by \autoref{prop:opt_lambda}. We first calculate that $cav(\hat{V}_{\lambda^*})=\frac{5}{2}$ and the process of concavification splits $\mu_0$ into $(\frac{1}{2},\frac{1}{4},\frac{1}{4}),(0,\frac{1}{2},\frac{1}{2})$ with probability $\frac{2}{3},\frac{1}{3}$ respectively. We denote this distribution over the posterior as $\tau^*$. We verify that $\mathcal{L}(\lambda_2,\tau^*)=\mathcal{L}(\lambda_3,\tau^*)=\frac{5}{2}<3=\mathcal{L}(\lambda_1,\tau^*)$. So by \autoref{prop:opt_lambda}, $\lambda^*$ is the worst subjective prior.

Moreover, we can show that under the binary-type cases, the credibility-gaining group exists for any $C$ unless MDMB collapses to CT. This is more general than \autoref{thm:credibility_gaining_group} because we do not impose a threshold on the budget. 

\begin{theorem}\label{thm:binary_credibility_gaining_group}
    Suppose the Sender has a binary type set. Under \autoref{ass:WeakDominated}, for any $C_2>C_1$, either $\mathcal{V}_{C_2}^*(\mu_0)>\mathcal{V}_{C_1}^*(\mu_0)$ or $\mathcal{V}^*_{C_2}(\mu_0)=\mathcal{V}^*_{0}(\mu_0)=\mathcal{V}^*_{CT}(\mu_0)$.
\end{theorem}
\begin{proof}[Proof of \autoref{thm:binary_credibility_gaining_group}]
Following the proof of \autoref{thm:credibility_gaining_group}, it suffices to observe that \( U_2 = \left\{ \mu \in \textbf{supp}\{p_{C_1}\} \,\middle|\, \sum_{\theta \in \Theta} \lambda_{C_2}(\theta) \frac{\mu(\theta)}{\mu_0(\theta)} = 0 \right\} \) contains at most one element. Therefore, by \autoref{cl:persuasion_to_CT}, any posterior that does not belong to the persuasion group must yield the same payoff as those within the persuasion group. This directly constructs a cheap talk equilibrium.

\end{proof}

\end{document}